\documentclass{IEEEtran}
\usepackage{cite}
\usepackage{hyperref}       
\hypersetup{hypertex=true,
colorlinks=true,
linkcolor=blue,
anchorcolor=blue,
citecolor=blue}
\usepackage{amsmath,amssymb,amsfonts}
\usepackage{algorithmic}
\usepackage{graphicx}
\usepackage{textcomp}
\usepackage{bm}
\usepackage{subfigure}
\usepackage{multirow}
\def\BibTeX{{\rm B\kern-.05em{\sc i\kern-.025em b}\kern-.08em
    T\kern-.1667em\lower.7ex\hbox{E}\kern-.125emX}}

\usepackage[table]{xcolor}

\begin{document}
\title{A Deep Learning Scheme of Electromagnetic Scattering From Scatterers With Incomplete Profiles}
\author{
  Ji-Yuan Wang, Xin-Yue Lou, Liang Zhang, Yun-Chuan Wang, and Xiao-Min Pan \IEEEmembership{Member, IEEE}
  \thanks{ This work was supported by NSFC under Grant 62171033. (Corresponding author: Xiao-Min Pan)}
  \thanks{ The authors are with the School of Cyberspace Science and Technology, Beijing Institute of Technology, Beijing 100081, China. (e-mail: xmpan@bit.edu.cn)}
}

\maketitle

\begin{abstract}
A deep learning scheme is proposed to solve the electromagnetic (EM) scattering problems where the profile of the dielectric scatterer of interest is incomplete.
As a compensation, a limited amount of scattering data is provided, which is in principle containing sufficient information associated with the missing part of the profile.  
The existing solvers can hardly realize the compensation if the known part of the profile and the scattering data are combined straightforwardly. 
On one hand, the well-developed forward solvers have no mechanism to accept the scattering data, which can recover the unknown part of the profile if properly used. 
On the other hand, the existing solvers for inverse problems cannot retrieve the complete profile with an acceptable accuracy from the limited amount of scattering data, even when the available part of the profile can be fed into the solvers.
This work aims to handle the difficulty. %
To this end, the EM forward scattering from an incompletely known dielectric scatterer is derived.
A scheme based on DL is then proposed where the forward and inverse scattering problems are solved simultaneously. 
Numerical experiments are conducted to demonstrate the performance of the proposed DL-based scheme for both two-dimensional (2-D) and three-dimensional (3-D) EM scattering problems.
\end{abstract}


\section{Introduction}
\label{sec:int}
\IEEEPARstart{A}{ccurate} modeling of electromagnetic (EM) scattering plays an important role in modern applications, including medical imaging, remote sensing, seismic exploration, and so on~\cite{10332205,6313880, BWeglein_2003,Abubakar_2009,doi:10.1190/1.3246586, bo-wenWideAngularSweeping2022}.
When the complete profile of a scatterer is available, including its geometry shape and constitutive parameters (e.g., permittivity and permeability), a number of well-developed methods can be employed to obtain its scattering. 
People often call such problem as a forward scattering problem, or a scattering problem/forward problem for short. 
When the profile of a scatterer is unknown, a set of methods are also available to retrieve the scatterer's profile when a sufficient amount of observation/measured scattering data can be accessed. 
The associated problem is called an inverse scattering problem~\cite{chen2018computational}.

High frequency-approximation methods and full-wave numerical ones are the commonly employed traditional numerical methods for forward problems. 
The former~\cite{7904648,99043,jin2006fast,10604770} is always suitable for electrically large problems.
The delivered accuracy is in essence hard to be controlled due to the underlying approximations employed.
The latter includes the finite difference time domain (FDTD) method~\cite{kunz1993finite,schneider2010understanding,teixeira2023finite}, the finite element method (FEM)~\cite{jin2015finite,pan2015prediction,jin2008finite} and the method of moments (MoM)~\cite{gibson2021method, pan2012solving,liu2022solution}.
They are more reliable in terms of accuracy at the expense of high computational costs.
In addition to the aforementioned traditional methods, the artificial intelligence-aided (AI-aided) techniques~\cite{shan2020study, 10516293, wang2022physics,10318053, 10014533, jywang2024phy,10590715, 10719669,
ma2020learning,xue2023deep,guo2021physics,10508749,10702499,9696232} attract more and more interests of researchers in recent years.

Similar to forward problems, methods for EM inverse scattering problems can also be divided into traditional and AI-aided ones.
Along the traditional technique line, several approaches based on linear approximation, such as Born approximation~\cite{54365}, back-propagation (BP)~\cite{belkebir2005superresolution} and Rytov approximation~\cite{Devaney81}, are able to provide accurate reconstruction results efficiently for scatterers with weakly scattering.
As scattering becomes stronger, these solvers suffer from inherent ill-posedness and nonlinearity. 
To address these obstacles, numerical solutions are often converted into optimization frameworks~\cite{10685032, 7406678, 6868244, xu2020fourier,yu2022fourier}. 
Along the AI-aided technical line, deep learning (DL) has also achieved impressive advance in EM inverse scattering problems~\cite{8709721,wei2018deep,pan2021phase,9214919,10058707,10380631,8747485,9509361,10225705,10454009}, benefiting from its intrinsic ability to handle nonlinearity. 
Many researches revealed that the AI-aided methods can often provide more accurate solutions than traditional ones.
However, the aforementioned methods do NOT cover all types of problems that exist in real world. 
For example, people may want to find scattering characteristic of a scatterer whose profile is only partly available. 
As a compensation to the missing part of the profile, a small amount of observation data can be accessed which are assumed able to provide the information the missing profile contained. 
Typically, the scattering data may be obtained by illuminating the scatterer by a single or a very few number of plane waves. %
It is obviously that existing methods both for forward and inverse problems are incapable of that problem.
On one hand, such a limited number of scattering data are indeed inadequate to retrieve the complete profile of the scatterer alone.
As a result, the well-developed solvers for inverse problems fail in solving it as an inverse problem even when modifications are done to integrate the known part of the profile into the solvers. 
On the other hand, without the complete profile, the existing forward solvers are incapable of that problem. 
This work aims to fill the gap associated with this type of problems by developing a DL-based learning scheme to make full use of the known part of the profile and the limited observation data.
The main contributions of this work are as follows.
\begin{enumerate}[]
  \item The formulation of the EM forward scattering from an incompletely known scatterer is derived.
  \item A DL-based learning scheme is developed according to the derived formulation. 
  \item A comparative study is conducted to demonstrate the performance of the proposed DL-based scheme for both two-dimensional (2-D) and three-dimensional (3-D) scatterers. 
\end{enumerate}

The rest of this work is as follows.
In Section~\ref{sec:Preliminaries}, notations and some basic concepts are given with respect to the EM scattering. 
In Section~\ref{sec:pm}, the proposed DL-based scheme for incompletely known scatterers is detailed after the associated formulation is derived.
In Section~\ref{sec:NR1}, numerical experiments are carried out to validate the proposed scheme.
Conclusions are given in Section~\ref{sec:Conclusion}.

\begin{figure}[!tp]
  \centering
  \includegraphics[width=0.28\textwidth]{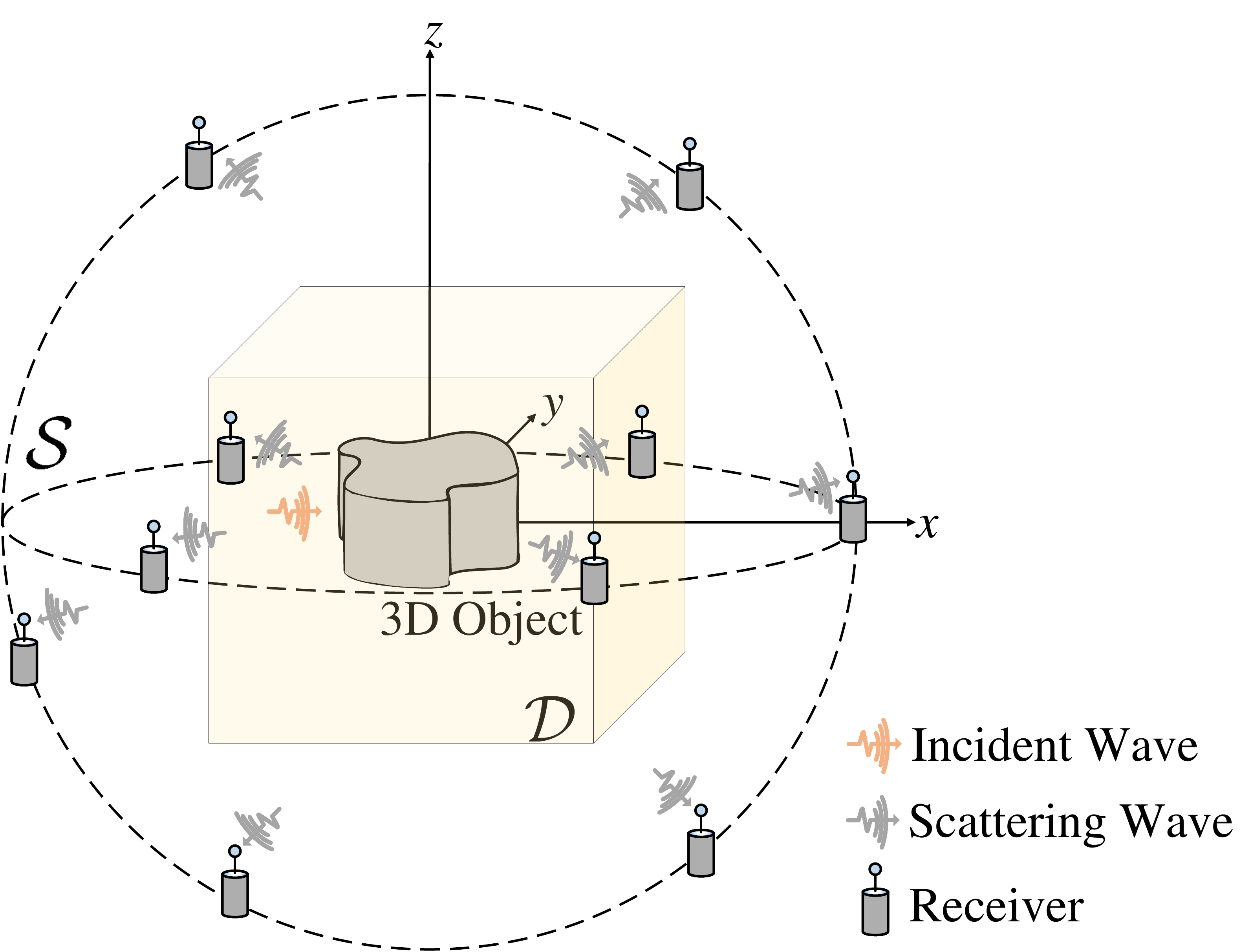}
\caption{Configuration of the forward/inverse scattering problem}
\label{fig:Setup}
\end{figure}

\begin{figure}[!tp]
\centering
\includegraphics[width=0.18\textwidth]{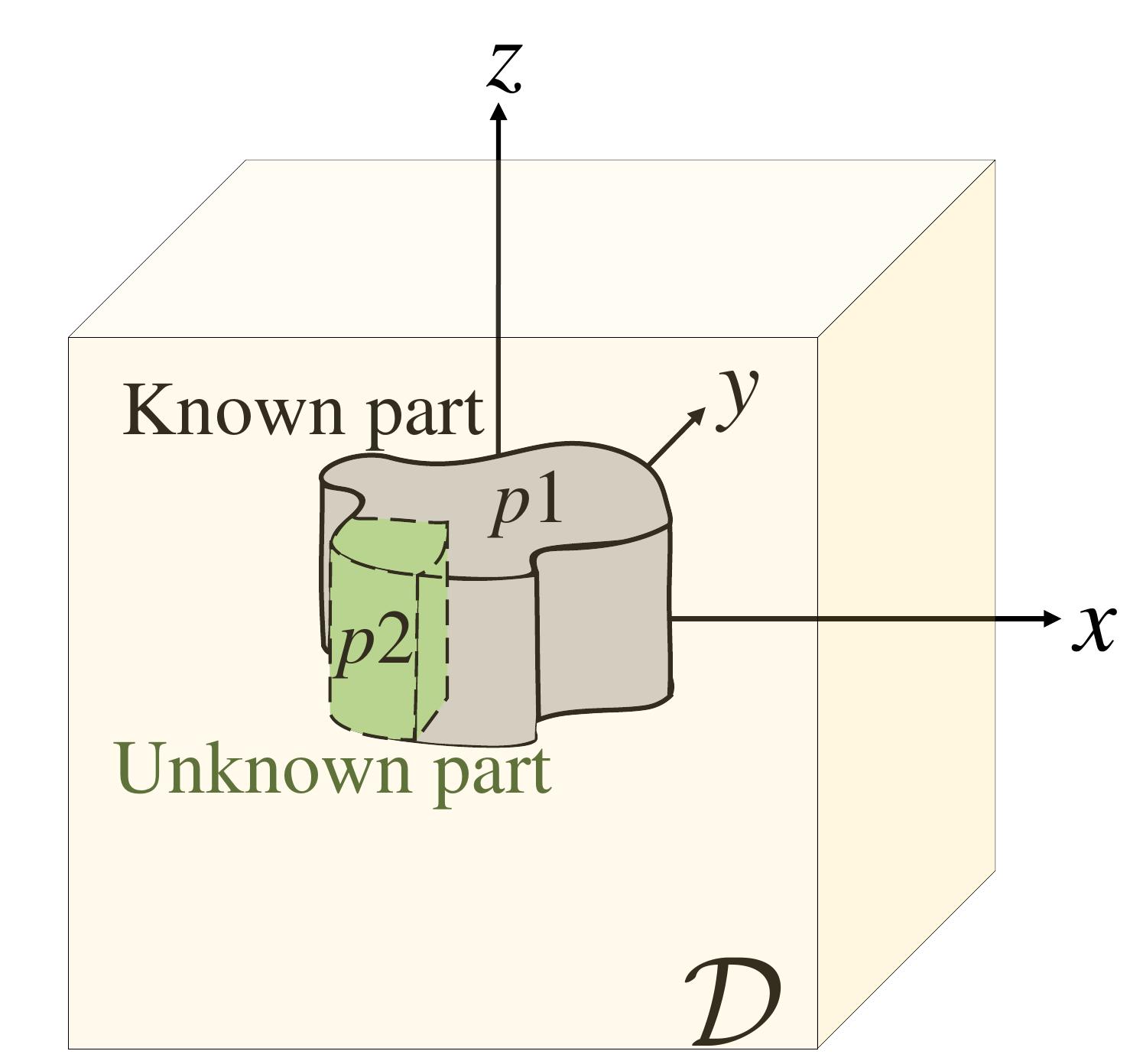}
\caption{An example of incomplete profile of a scatterer. }
\label{fig:Setup_incomplete}
\end{figure}

\section{Preliminaries}
\label{sec:Preliminaries}

Our discussions are restricted to the frequency domain and the time convention of $e^{\jmath \omega t}$ is omitted, where $\jmath = \sqrt {-1}$ refers to the imaginary unit and $\omega$ is the angular frequency.
\subsection{EM Forward and Inverse Scattering Problems}
\label{secL:efs}
Under the Cartesian coordinate system, assume a 3-D nonmagnetic dielectric scatterer presented in the free space is enclosed by $\mathcal{D} \in \mathbb{R}^{3}$, the permeability of the scatterer equals to that of the free space $\mu_0$ and its permittivity at $\mathbf{r}=\left(x, y, z\right)$ is given by
\begin{equation}
	\label{eq:permittivity}
	\epsilon \left( {\mathbf{r}} \right) = {\epsilon _0}{\epsilon _r}\left( {\mathbf{r}} \right) + \jmath\frac{{\sigma \left( {\mathbf{r}} \right)}}{\omega },
\end{equation}
where ${\epsilon _0}$ is the permittivity of the free space, ${\epsilon _r}\left( {\mathbf{r}} \right)$ is the relative permittivity at $\mathbf{r}$, and ${\sigma \left( {\mathbf{r}} \right)}$ is the conductivity at $\mathbf{r}$.
A time-harmonic plane wave illuminates on domain $\mathcal{D}$, where $\mathbb{E}^{\text{inc}}(\mathbf{r})$ denotes the incident field at $\mathbf{r}$. 
The configuration of the associated 3-D EM forward scattering problem is shown in Fig.~\ref{fig:Setup}.
The 2-D case can be similarly defined, which is not presented to save space. 
The total field within $\mathcal{D}$ can be written in the form of the volume integral equation (VIE)~\cite{chew2022integral,chew1999waves,10103814, Polimeridis2014StableFS},
\begin{equation}
  \label{eq:EFIF}
  \begin{aligned}
    \mathbb{E}^\text{tot}\left( {\mathbf{r}} \right) = &\mathbb{E}^{{\text{inc}}}\left( {\mathbf{r}} \right) \\
    &+ k_0^2\int_\mathcal{D} G\left( {{\mathbf{r}},{\mathbf{r^{\prime}}}} \right)\chi \left( {\mathbf{r}^{\prime}} \right)\mathbb{E}^\text{tot}\left( {{\mathbf{r^{\prime}}}} \right) d{\mathbf{r^{\prime}}}, 
    \quad {\mathbf{r}} \in \mathcal{D},
    \end{aligned}
\end{equation}
where $k_0$=$\omega\sqrt{\epsilon_0\mu_0}$ is the wavenumber in free space,
$\mathbf{r}$ and $\mathbf{r}^\prime$ denote the positions of the observation and source point, respectively. 
In Eq.~\eqref{eq:EFIF}, $G\left( {{\mathbf{r}},{\mathbf{r'}}} \right)$ is the 3-D Green's function in free space and $\chi \left( {\mathbf{r^{\prime}}} \right)$ is the contrast function, which can be written as,
  \begin{equation}
    \label{eq:G_3d}
    G\left( {{\mathbf{r}},{\mathbf{r'}}} \right) = \frac{{{e^{ - \jmath{k_0}\left| {{\mathbf{r}} - {\mathbf{r'}}} \right|}}}}{{4\pi \left| {{\mathbf{r}} - {\mathbf{r'}}} \right|}} ~\text{and}~ 
	  \chi \left( {\mathbf{r^{\prime}}} \right) = \frac{{\epsilon \left( {\mathbf{r^{\prime}}} \right) - {\epsilon _0}}}{{{\epsilon _0}}}.
  \end{equation}

Assuming  there is a sufficiently large surface $\mathcal{S}$ in free space that encloses domain $\mathcal{D}$, $N_s$ receivers are located to measure the scattering field that can be theoretically calculated by
\begin{equation}
  \label{eq:es}
  \begin{aligned}
    \mathbb{E}^\text{sca}\left( {\mathbf{r}} \right) = k_0^2\int_\mathcal{D} {G\left( {{\mathbf{r}},{\mathbf{r^{\prime}}}} \right)\chi \left( {\mathbf{r^{\prime}}} \right)\mathbb{E}^\text{tot}\left( {{\mathbf{r^{\prime}}}} \right)d{\mathbf{r'}}}, \quad \mathbf{r} \in \mathcal{S}.
  \end{aligned}
\end{equation}
 
It should be noted that since the Cartesian coordinate is employed in our work, the corresponding parameters and fields can have $x$-, $y$- and $z$-components for 3-D scatterers.
For example, $\chi$ has $\chi_x$, $\chi_y$ and $\chi_z$ three components.   
We would like to omit these subscripts when no confusion will arise.  
Without loss of generality, MoM can solve the forward scattering problem Eq.~\eqref{eq:EFIF} if the complete profile $\chi$ is available.
In MoM, the domain $\mathcal{D}$ is discretized into totally $M=M_1\times M_2 \times M_3$ voxels, where $M_1$, $M_2$ and $M_3$ are the numbers of segments along $x$-, $y$- and $z$-axes.
Discretizing Eq.~\eqref{eq:EFIF} by the pulse basis function and the delta testing function, we have,
\begin{equation}
  \label{eq:dis_j}
  \mathbf{E}^\text{tot} = \mathbf{E}^{{\text{inc}}} + {{\mathbf{ { G}}}_\mathcal{D}} \cdot {{\bm{\chi}}} \cdot \mathbf{E}^\text{tot},
\end{equation} %
where $\mathbf{E}^\text{tot}=[\mathbf{E}_x^\text{tot}, \mathbf{E}_y^\text{tot}, \mathbf{E}_z^\text{tot}]^T\in \mathbb{C}^{3M\times1}$ is the vector of the discrete total field within domain $\mathcal{D}$,
$\bm{\chi}\in \mathbb{C}^{3M\times3M}$ is the diagonal matrix storing three components of $\chi$ for each voxel, 
${{{\mathbf{ { G}}}}_\mathcal{D}} \in \mathbb{C}^{3M\times3M}$ is the matrix form of $G\left( {{\mathbf{r}},{\mathbf{r'}}} \right)$ with respect to domain $\mathcal{D}$.
Equation~\eqref{eq:es} can be similarly discretized as,
\begin{equation}
  \label{eq:dis_es}
  \mathbf{E}^\text{sca} = {\mathbf{ { G}}}_{\mathcal{S}} \cdot {{\bm{ {\chi} }}} \cdot \mathbf{E}^\text{tot},
\end{equation}
where $\mathbf{E}^\text{sca}=[\mathbf{E}_x^\text{sca}, \mathbf{E}_y^\text{sca}, \mathbf{E}_z^\text{sca}]^T\in \mathbb{C}^{3N_s\times1}$ is the vector of the discrete scattering fields on $\mathcal{S}$,  and ${{\mathbf{ { G}}}_\mathcal{S}} \in \mathbb{C}^{3N_s\times3M}$ is the matrix form of $G\left( {{\mathbf{r}},{\mathbf{r'}}} \right)$ with respect to $\mathcal{S}$.

An EM inverse scattering problem is concerned with retrieving the profile of an unknown scatterer, such as its geometry and constitutive parameters, from a set of scattering fields ${\mathbf{E}}^\text{sca}$'s.
In the framework of the EM inverse scattering problem, Eqs.~\eqref{eq:dis_j} and~\eqref{eq:dis_es} are referred to as the state equation and the data equation, respectively.   
The unknown scatterer is retrieved by minimizing an objective function that is a linear combination of normalized mismatches in the data equation and the state equation.

\subsection{Problem To Be Solved}
\label{sec:ps}
The problem to be solved in this work is depicted in Fig.~\ref{fig:Setup_incomplete}, where the profile of the scatterer is divided into a known region $p1$ and an unknown part $p2$. 
Assuming the diagonal matrix of the contrast for the known part of the scatterer is represented by ${{{\bm{{\chi}}}}^{p1}}$ while that for the unknown part by ${{\bm{\chi}}^{p2}}$, we have, 
\begin{equation}
  \label{eq:xxx}
  \bm{\chi} = \bm{\chi}^{p1}+\bm{\chi}^{p2}.
\end{equation}
Along with $\bm{\chi}^{p1}$, we have access to the scattering data with respect to a limited number of plane wave illuminations. 
The incident angle can be randomly chosen in this work.
Our goal is to predict scattering fields from the scatterer under any incident and any observation angles. %

\begin{figure*}[!ht]
  \centering
  \includegraphics[width=0.8\textwidth]{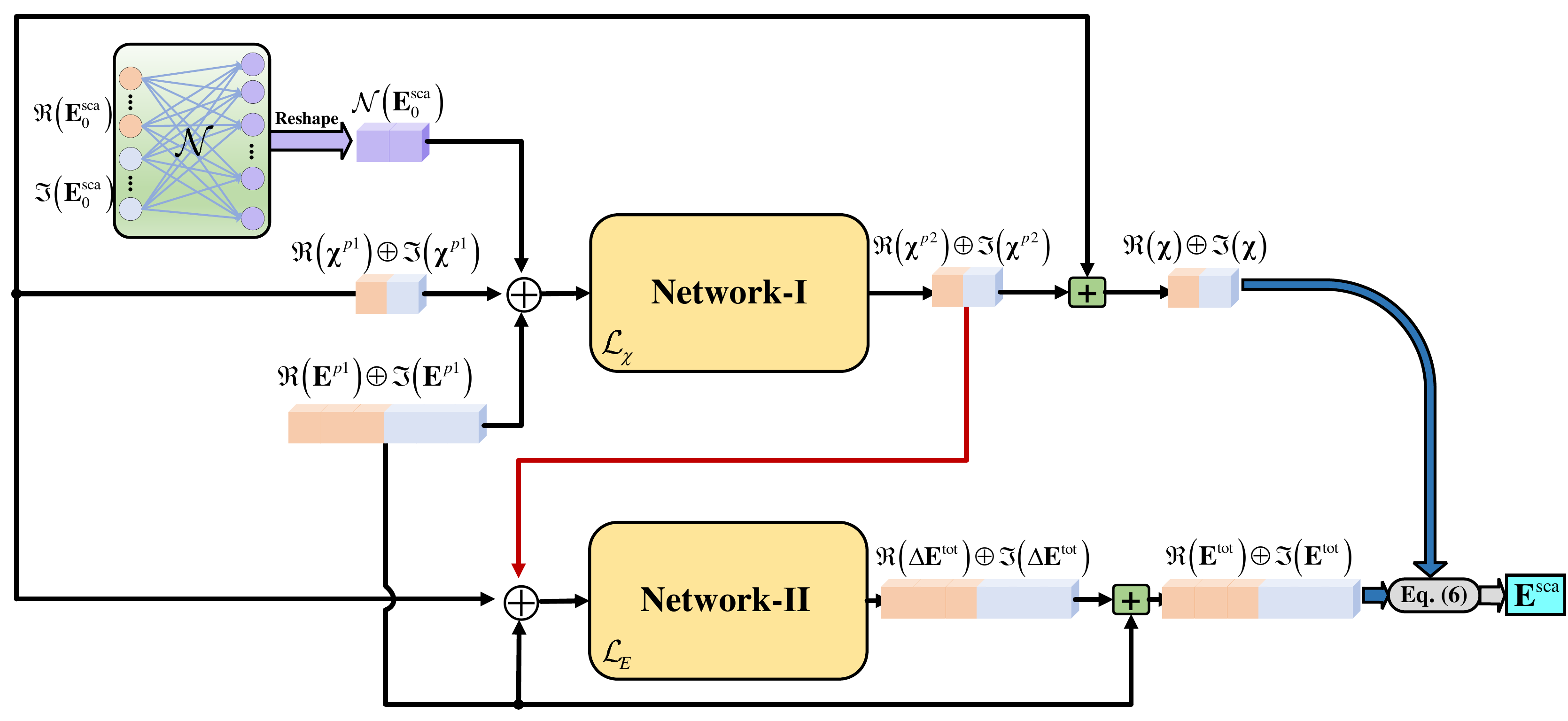}
  \caption{The overflow of the developed DL-based learning scheme.
  $\oplus$ means to put the matrices on both sides of the symbol into the two channels of the tensor.
  $\Re(\cdot)$ and $\Im(\cdot)$ denote to take the real and imaginary part of the corresponding tensor, respectively.}
\label{fig:scheme}
\end{figure*}

\begin{figure}[]
  \centering
  \includegraphics[width=0.45\textwidth]{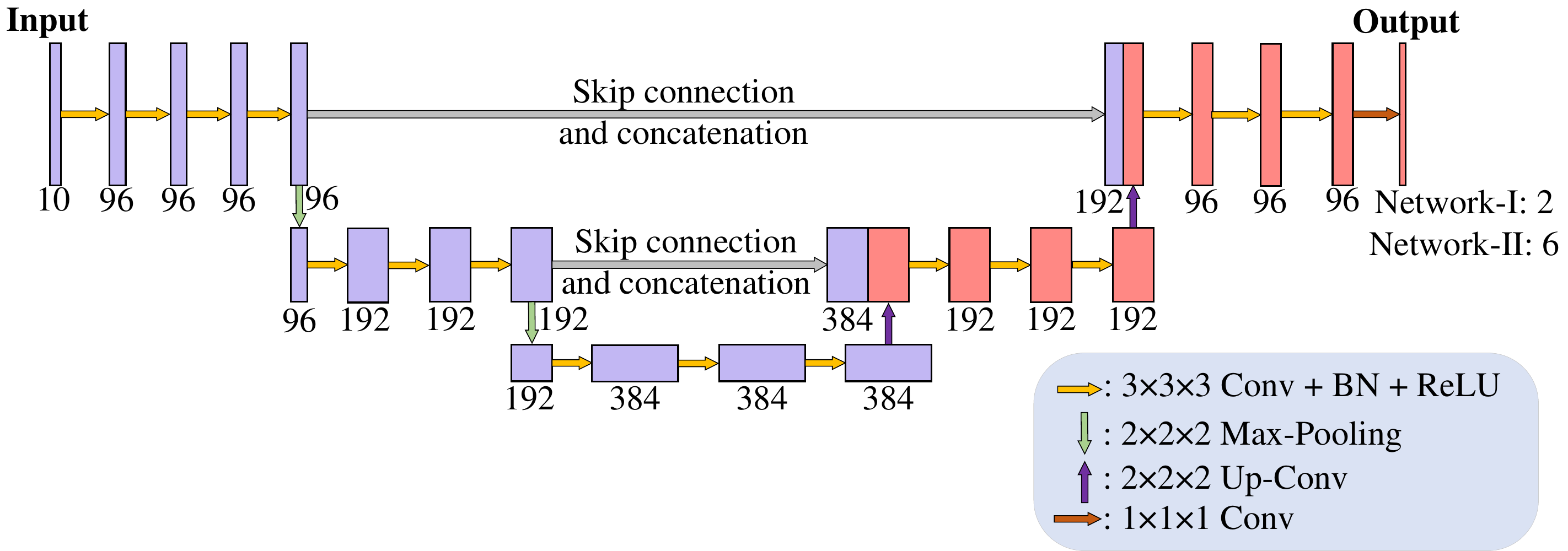}
\caption{The sketch of the employed 3-D U-Net.}
\label{fig:unet}
\end{figure}

\section{Proposed Method}
\label{sec:pm}
The formulation of scattering fields involving both $\bm{\chi}^{p1}$ and $\bm{\chi}^{p2}$ is deduced in this section. 
A DL-based scheme that can make full use of the known part of the profile and the limited observation data is then developed accordingly.

\subsection{Formulation of Scattering Fields}
\label{sec:Formulation}

With Eq.~\eqref{eq:xxx}, Eq.~\eqref{eq:dis_j} can be rewritten as
\begin{equation}
	\label{eq:dis_e_r}
	{\mathbf{E}^\text{tot}} = \mathbf{E}^{{\text{inc}}} + {{\mathbf{ { G}}}_\mathcal{D}} \cdot \left( {\bm{\chi}^{p1} + {{\bm{\chi}}^{p2}}} \right) \cdot {\mathbf{E}^\text{tot}}.
\end{equation}
By a simple deduction, we can get
\begin{equation}
	\label{eq:dis_e_rr}
	{\mathbf{E}^\text{tot}} = {\left( {{\mathbf{ { I}}} - {\mathbf{ { A}}} \cdot\bm{\chi}^{p2}} \right)^{ - 1}} \cdot {\mathbf{E}^{p1}},
\end{equation}
where 
\begin{equation}
  \label{eq:ep1z}
  \begin{split}
\mathbf{ { A}} &= {\left( {{\mathbf{ { I}}} - {{\mathbf{ { G}}}_\mathcal{D}} \cdot \bm{\chi}^{p1}} \right)^{ - 1}} \cdot {{\mathbf{ { G}}}_\mathcal{D}} \\
%
%
  {\mathbf{E}^{p1}} &= {\left( {{\mathbf{ { I}}} - {{\mathbf{ { G}}}_\mathcal{D}} \cdot \bm{\chi}^{p1}} \right)^{ - 1}} \cdot \mathbf{E}^{{\text{inc}}}
  \end{split}.
\end{equation}
Here, ${\mathbf{E}^{p1}}$ represents the total field from part $p1$ only, or to say, the field without the presence of $p2$.
Thus, the contribution from part $p2$ can be defined as
\begin{equation}
	\label{eq:dis_e_con}
  \begin{split}
	\Delta {\mathbf{E}^\text{tot}}&\mathop= \limits^\Delta {\mathbf{E}^\text{tot}} - {\mathbf{E}^{p1}} = \left[ {{{\left( {{\mathbf{ { I}}} - {\mathbf{ { A}}} \cdot \bm{\chi}^{p2}} \right)}^{ - 1}} - {\mathbf{ { I}}}} \right] \cdot {\mathbf{E}^{p1}}\\
  & = \mathcal{L}_E\left(\bm{\chi}^{p1}, \bm{\chi}^{p2}, {\mathbf{E}^{p1}}\right).
  \end{split}
\end{equation}
In Eq.~\eqref{eq:dis_e_con}, the nonlinear discrete operator $\mathcal{L}_E$ is introduced for brevity. 
It should be noted that $\Delta {\mathbf{E}^\text{tot}}$ contains the interaction between the fields within $p2$ and $p1$. 
As a result, $\Delta {\mathbf{E}^\text{tot}}$ cannot be simply computed by ${\left( {{\mathbf{ { I}}} - {{\mathbf{ { G}}}_\mathcal{D}} \cdot \bm{\chi}^{p2}} \right)^{ - 1}} \cdot \mathbf{E}^{{\text{inc}}}$ even if $\bm{\chi}^{p2}$ is retrieved.  
%

Substituting Eqs.~\eqref{eq:xxx} and~\eqref{eq:dis_e_rr} into Eq.~\eqref{eq:dis_es} yields, 
\begin{equation}
	\label{eq:xi_esm2}
	{\mathbf{E}^\text{sca}}={{\mathbf{ { G}}}_\mathcal{S}} \cdot \left(\bm{\chi}^{p1}+\bm{\chi}^{p2}\right) \cdot {\left( {{\mathbf{ { I}}} - {\mathbf{ { A}}} \cdot \bm{\chi}^{p2}} \right)^{ - 1}} \cdot {\mathbf{E}^{p1}}.
\end{equation}
From Eq.~\eqref{eq:xi_esm2}, the following equation can be reached,
\begin{equation}
	\label{eq:lcon}
  \begin{split}
	{\bm{\chi}}^{p2} &= {\left( {{\mathbf{ { I}}} + \frac{{{\mathbf{ { G}}}_\mathcal{S}^\dag  \cdot {{\mathbf{E}}^{{\text{sca}}}} \cdot {{\left( {{{\mathbf{E}}^{p1}}} \right)}^H} \cdot {\mathbf{ { A}}}}}{{{{\left( {{{\mathbf{E}}^{p1}}} \right)}^H} \cdot {{\mathbf{E}}^{p1}}}}} \right)^{ - 1}} \\
  &\cdot \left( {\frac{{{\mathbf{ { G}}}_\mathcal{S}^\dag  \cdot {{\mathbf{E}}^{{\text{sca}}}} \cdot {{\left( {{{\mathbf{E}}^{p1}}} \right)}^H}}}{{{{\left( {{{\mathbf{E}}^{p1}}} \right)}^H} \cdot {{\mathbf{E}}^{p1}}}} - {\bm{\chi}}^{p1}} \right) \\
  &= \mathcal{L}_{\chi}\left(\bm{\chi}^{p1},{\mathbf{E}^\text{sca}},{\mathbf{E}^{p1}}\right),
  \end{split}
\end{equation}
where $H$ denotes the conjugate transpose operation and $\dag$ means the pseudo-inverse operation.
Again, for brevity, an operator similar to that in Eq.~\eqref{eq:dis_e_con} is defined in Eq.~\eqref{eq:lcon}. 
Clearly, once $\bm{\chi}^{p2}$ is retrieved, $\Delta{\mathbf{E}^\text{tot}}$ and thus ${\mathbf{E}^\text{tot}}$ can be obtained from the complete profile of the scatterer. 
That is to say, scattering fields under any incident and any observation direction can be calculated through Eq.~\eqref{eq:dis_es}.
In principle, $\bm{\chi}^{p2}$ can be retrieved according to the amount of scattering data available with the aid of the known part of the profile $\bm{{ \chi}}^{p1}$ through Eq.~\eqref{eq:lcon}. 
The observation data, $\mathbf{E}^\text{sca}$, is usually obtained by placing a sufficiently large number of transiting and receiving antennas around the scatterer for a typical inverse problem. 
However, in this work, the number of transiting antennas is as small as 1 because we have $\bm{\chi}^{p1}$. %
In what follows, $\mathbf{E}^\text{sca}_0$ is employed instead of $\mathbf{E}^\text{sca}$ to emphasis the amount of the data is limited in our work.
It is important to make good use of $\bm{\chi}^{p1}$, ${\mathbf{{ E}}}^\text{sca}_0$ and ${\mathbf{{ E}}}^{p1}$ to retrieve $\bm{\chi}^{p2}$, as shown in Eq.~\eqref{eq:lcon}.
As will be numerically revealed by the results in the last column in Fig.~\ref{fig:pred_result}, a slightly modified SOM~\cite{chen2009subspace}\footnote{The SOM is an advanced traditional inverse problem solver.} struggles to resolve $\bm{\chi}^{p2}$ under a single or a very few number of incident waves even if $\bm{\chi}^{p1}$ are utilized to construct the initial values for these numerical algorithms.
Besides the inherent ill-posedness and nonlinearity of the inverse problems, one reason lies in that the SOM does not consider the coupling between $p1$ and $p2$ as exhibited in Eq.~\eqref{eq:dis_e_con}. 
Conversely, Eq.~\eqref{eq:lcon} makes good use of the coupling between $p1$ and $p2$ by explicitly utilizing $\mathbf{E}^{p1}$ as well as $\bm{\chi}^{p1}$ as input. 
If modified to exploit the coupling properly, we believe, traditional solvers including the SOM can perform well. 
Of course, the effort associated with the modification may depend heavily on the inherent mechanism of the solver to be modified. 
Recent advances indicates that DL techniques can mitigate the ill-posedness and nonlinearity of the original inverse problems.
Encouraged by this, this work develops a DL-based scheme to realized the solution process associated with Eqs.~\eqref{eq:lcon} and~\eqref{eq:dis_e_con}.

\subsection{Proposed DL-based Scheme}
\label{sec:DL SCHEME}
As shown by Fig.~\ref{fig:scheme}, the propose scheme consists of two sub-networks, namely, Network-I and Network-II. 
In the figure, $\mathbf{C1}\oplus \mathbf{C2}$ means the operation to put the matrices $\mathbf{C1}$ and $\mathbf{C2}$ into two channels, $\Re(\cdot)$ and $\Im(\cdot)$ denote operators to take the real and imaginary parts of the corresponding tensor.
Since the inputs and outputs of the proposed DL scheme usually align with the spatial distribution of physical parameters, the discrete parameters in Sections~\ref{sec:Preliminaries} and~\ref{sec:Formulation} should be converted into their corresponding spatial distribution representations. 
For example, $\bm{\chi}^{p1}$, a $3M \times 3M$ matrix, should be a $3\times M_1 \times M_2 \times M_3$ tensor for a 3-D scatterer. 
Fortunately, it is quite obvious when such a conversion is required.  
To make our notations simple, this work doesn't distinguish such difference arising from the conversion in the following discussion. 

Network-I is aimed to approximate operator $\mathcal{L}_{\chi}$ in Eq.~\eqref{eq:lcon}. 
Its input contains $\bm{\chi}^{p1}$, $\mathcal{N}\left(\mathbf{ {E}}^{\text{sca}}_\text{0}\right)$, and ${\mathbf{E}^{p1}}$, where $\mathcal{N}$ consists of a linear fully connected layer followed by a rectified linear unit (ReLU) activation function. 
Its output is $\bm{\chi}^{p2}$.
Taking a 3-D scatterers as an example, as shown in Fig.~\ref{fig:scheme}, $\mathcal{N}$ takes $\mathbf{ { E}}^{\text{sca}}_\text{0}$ as input and reshapes output into a tensor with the size of $2\times M_1\times M_2 \times M_3$.

Network-II is to approximate operator $\mathcal{L}_{E}$ in Eq.~\eqref{eq:dis_e_con}.
The input of the network is a concatenation of $\bm{\chi}^{p1}$, ${\mathbf{E}^{p1}}$ and $\bm{\chi}^{p2}$ predicted by Network-I.
The associated output is $\Delta{\mathbf{E}^\text{tot}}$.
The predicted $\Delta{\mathbf{E}^\text{tot}}$ is employed to generate the scattering field, which, in turn, is employed to evaluate the accuracy of loss of the whole network.
In short, after $\bm{\chi}^\text{p2}$ is retrieved, the scattering field at any observation point
can be calculated through Eq.~\eqref{eq:dis_es}.

The loss function of Network-I, denoted by $loss_\text{I}$, is a function of the predicted complete contrast $\bm{\chi}=\bm{\chi}^{p1}+\bm{\chi}^{p2}$. 
That of Network-II, denoted by $loss_\text{II}$, is a function of the predicted total field ${\mathbf{E}^\text{tot}}={\mathbf{E}^{p1}}+{\Delta\mathbf{E}^\text{tot}}$. 
The mean square error (MSE) between the predicted results and their ground true counterparts is employed to define the loss functions. 
More specifically, 
\begin{equation}
  \label{eq:loss_xi}
  \begin{aligned}
    &loss_\text{I} = \frac{1}{B}\sum\limits_{b = 1}^B {\frac{{{{\left\|\bm{\chi}^\text{Pred}_b - {\bm{\chi}^\text{Label}_b} \right\|}_F}}}{M}}; \\
    &loss_\text{II} = \frac{1}{B}\sum\limits_{b = 1}^B {\frac{{{{\left\| \mathbf{E}_b^{{\text{tot, Pred}}} -  \mathbf{E}_b^{{\text{tot, Label}}}  \right\|}_F}}}{M}},
  \end{aligned}
\end{equation}
where $B$ is the batch size, $\left\|\cdot\right\|_F$ denotes the Frobenius norm of a matrix, $\bm{\chi}_b^\text{Pred}$ and $\mathbf{E}_b^{{\text{tot, Pred}}}$ represent the predicted results for the $b$-th scatterer through the network, $\bm{\chi}_b^\text{Label}$ and $\mathbf{E}_b^{{\text{tot, Label}}}$ are the corresponding labels.
The total loss of the cascaded DL-based scheme can be calculated by summing the loss of Network-I and Network-II, which can be written as, 
\begin{equation}
	\label{eq:loss}
  \begin{aligned}
    loss_\text{Total} = \beta_1loss_\text{I}+\beta_2loss_\text{II},
  \end{aligned}
  \end{equation}
where $\beta_1$ and $\beta_2$ are the prescribed hyperparameters to balance the two losses. 
This work simply sets $\beta_1=\beta_2=1/2$.

Except for the input-output pair of Network-I and Network-II discussed here, several alternative choices are investigated in Section~\ref{sec:fv2d}.


\subsection{Backbone Network}
\label{sec:ns}
The backbone is a commonly used end-to-end U-Net~\cite{ronneberger2015u, cciccek20163d}, which consists of a contracting path and an expansive path.
Figure~\ref{fig:unet} gives a sketch of the employed 3-D U-Net~\footnote{A 2-D U-Net is employed for 2-D scatterers, which shares the same structure to the 3-D one, except that the convolutional kernels are reduced to 2-D dimensions.} for 3-D scatterers, where the number of channels is presented below each layer.
The U-Net contains $3 \times 3 \times 3$ convolutions with stride 1 and padding 1, each followed by a batch normalization and a ReLU activation function. 
Additionally, a $2 \times 2 \times 2$ max-pooling operation with stride 2 is employed for downsampling along with the operation to double the number of channels. 
To reconstruct features from the corresponding latent representations, a $2 \times 2 \times 2$ up-convolution that halves the number of channels is utilized. 
A $1\times 1 \times 1$ convolution with stride 1 is employed to output the reconstructed picture whose map is with a size identical to that of the input map in our application.
Moreover, the skip connection and concatenation enable the fusion of features in the contracting and expansive paths.

\begin{figure}[!t]
  \centering
  \subfigure[A 2-D case.]{
    \includegraphics[width=0.45\textwidth]{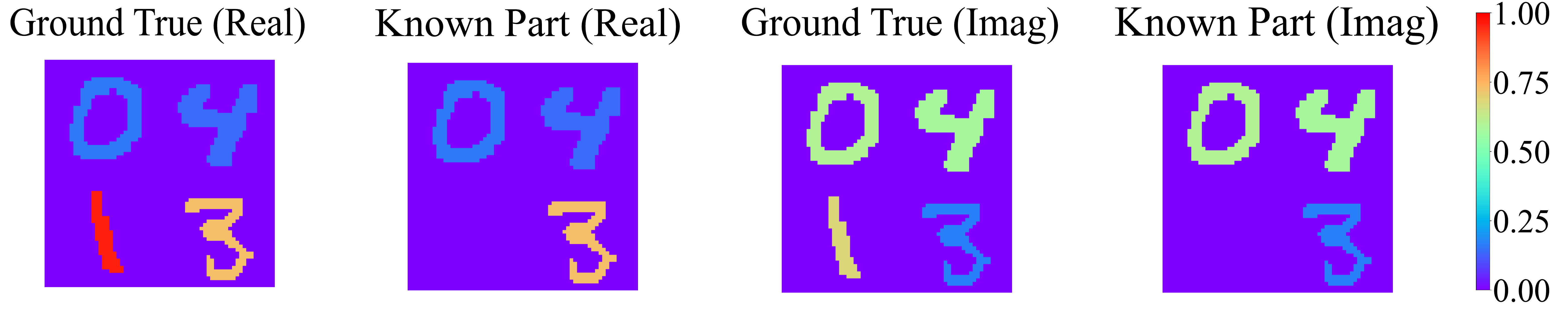}
    \label{fig:esc_2d}}
    \subfigure[A 3-D case.]{
      \includegraphics[width=0.45\textwidth]{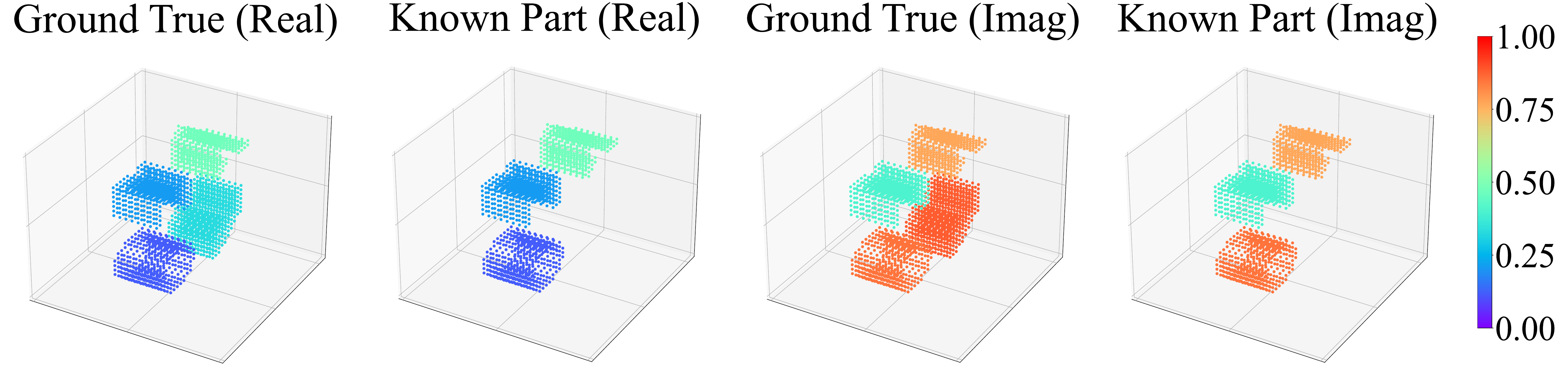}
      \label{fig:esc}}
\caption{Examples of samples contained in the MNIST training set.
The first and second columns show the real part of the complete and incomplete profiles, respectively.
The third and forth columns present the imaginary part of the complete and incomplete profiles, respectively.
 }
\end{figure}

\begin{table*}[!thpb]
  \begin{center}
    \caption{MREs for different input-output pairs of the DL-based scheme in the 2-D case.
    }
    \label{tab:di_test3}
    \setlength{\tabcolsep}{1mm}
    \begin{tabular}{|c|cc|cc|c|c|c|}
      \hline
      \multirow{2}{*}{Pair No.}&\multicolumn{2}{|c|}{Network-I}   & \multicolumn{2}{c|}{Network-II}     &\multicolumn{3}{c|}{MRE (\%)}  
      \\ \cline{2-8}
                               &Input & Output & Input & Output                                         & $~~\bm{\chi}^\text{Pred}~~$  &${\mathbf{E}^\text{tot, Pred}}$      &   $\mathbf{E}^\text{sca, Pred}$                         \\ \hline
     I &\multicolumn{1}{|c}{$\mathcal{N}\left(\mathbf{E}^\text{sca}_0 \right), \mathbf{E}^{\text{inc}}$}          &      $\bm{\chi}$      & \multicolumn{1}{c}{$\bm{\chi}, \mathbf{E}^{\text{inc}}$}    &   $\mathbf{E}^{\text{tot}}$     &      20.63            &               6.97              &       13.24            \\ 
     II &\multicolumn{1}{|c}{$\bm{\chi}^{p1}, \mathbf{E}^{p1}$}          &      $\bm{\chi}^{p2}$      & \multicolumn{1}{c}{$\bm{\chi}^{p1},\bm{\chi}^{p2}, \mathbf{E}^{p1}$}    &   $\Delta\mathbf{E}^{\text{tot}}$    &       15.27       &           14.38              &         29.38          \\ 
     \rowcolor{lime}
     III (Best) &\multicolumn{1}{|c}{$\bm{\chi}^{p1}, \mathcal{N}\left(\mathbf{E}^\text{sca}_0 \right),\mathbf{E}^{p1}$}          &      $\bm{\chi}^{p2}$      & \multicolumn{1}{c}{$\bm{\chi}^{p1}, \bm{\chi}^{p2}, \mathbf{E}^{p1}$}    &   $\Delta\mathbf{E}^{\text{tot}}$         &  {\bf 9.32}          &    {\bf2.68}        &       {\bf5.75}              \\ 
     %
     %
     IV &\multicolumn{1}{|c}{$\bm{\chi}^{p1}, \mathcal{N}\left(\mathbf{E}^\text{sca}_0 \right),\mathbf{E}^{\text{inc}}$}          &      $\bm{\chi}^{p2}$      & \multicolumn{1}{c}{$\bm{\chi}, \mathbf{E}^{\text{inc}}$}    &   $\mathbf{E}^{\text{tot}}$         &  9.88 &   3.18    &    7.54      \\     \hline        
     %
     %
      \end{tabular}
  \end{center}
\end{table*}

\section{Numerical Experiments}
\label{sec:NR1}
Numerical experiments are conducted by a server with a NVIDIA RTX 3090 GPU and a Xeon 6139 CPU on isotropic scatterers including dielectric metasurfaces~\cite{an2021deepconvolutionalneuralnetworks}, as well as MNIST~\cite{726791} and EMNIST~\cite{7966217} objects, to validate the performance of the proposed DL-based learning scheme.

The proposed scheme is implemented by Pytorch~\cite{ketkar2021introduction}.
In the training stage, the learning rates of both Network-I and Network-II are initialized to 0.001 and are halved per 50 epochs. 
The total number of epochs is set to 300 with a batch size $B=30$. 
The adaptive moment estimation method (Adam)~\cite{kingma2014adam} is employed to minimize the loss function. 
In the testing stage, to quantify the prediction accuracy, the mean relative error (MRE) is used,
\begin{equation}
  \label{eq:re_q}
  \begin{aligned}
    &{\text{MRE}} = \sum\limits_{q = 1}^Q {{\text{R}}{{\text{E}}_q}}, \\
    &{\text{R}}{{\text{E}}_q} = \frac{{{{\left\| {{\mathbf{A}^{{\text{Pred}}}_q} - {\mathbf{A}^{{\text{Label}}}_q}} \right\|}_F}}}{{\left\| {{\mathbf{A}^{{\text{Label}}}_q}} \right\|{}_F}}. \\
  \end{aligned}
\end{equation}
where $\text{RE}_q$ is the relative error (RE) for scatterer $q$, ${\mathbf{A}^\text{Pred}_q}=\bm{\chi}^\text{Pred} / \mathbf{E}^\text{tot, Pred} / \mathbf{E}^\text{sca, Pred} $ is the predicted result by the DL-based scheme for the $q$-th scatterer, ${\mathbf{A}^{{\text{Label}}}_q}=\bm{\chi}^\text{Label} / \mathbf{E}^\text{tot, Label} / \mathbf{E}^\text{sca, Label}$ is the corresponding label by MoM, and $Q$ is the total number of testing samples.

\begin{figure*}[!htpb]
  \centering
  \includegraphics[width=0.98\textwidth]{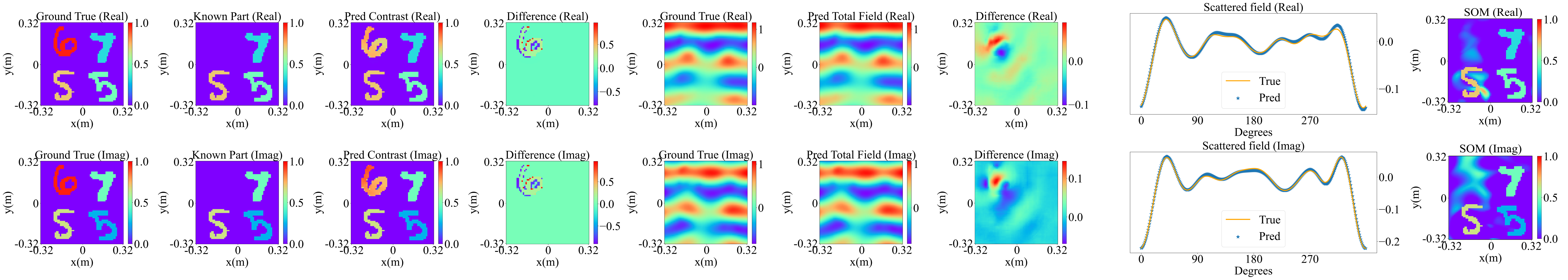}
\caption{2-D scatterers: predicted results of the proposed DL-based scheme as well as the SOM.
The first column shows the real and imaginary parts of $\bm{\chi}^\text{Label}$. 
The second column shows the real and imaginary parts of the known part.
The third and sixth columns display the real and imaginary parts of $\bm{\chi}^\text{Pred}$ and $\mathbf{E}^\text{tot, Pred}$ by the proposed DL-based scheme.
The forth and seventh columns show the real and imaginary parts of the differences between the predicted and the reference results. 
The fifth column presents the real and imaginary parts of $\mathbf{E}^\text{tot, Label}$ obtained by the MoM.  
The eighth column gives the real and imaginary parts of $\mathbf{E}^\text{sca, Pred}$ within [$0^\circ$, $360^\circ$] with respect to the $+x$-axis. 
The last column shows the real and imaginary parts of $\bm{\chi}^\text{Pred}$ by the SOM.
}
\label{fig:pred_result}
\end{figure*}
\begin{figure*}[!tpb]
  \centering
  \includegraphics[width=0.98\textwidth]{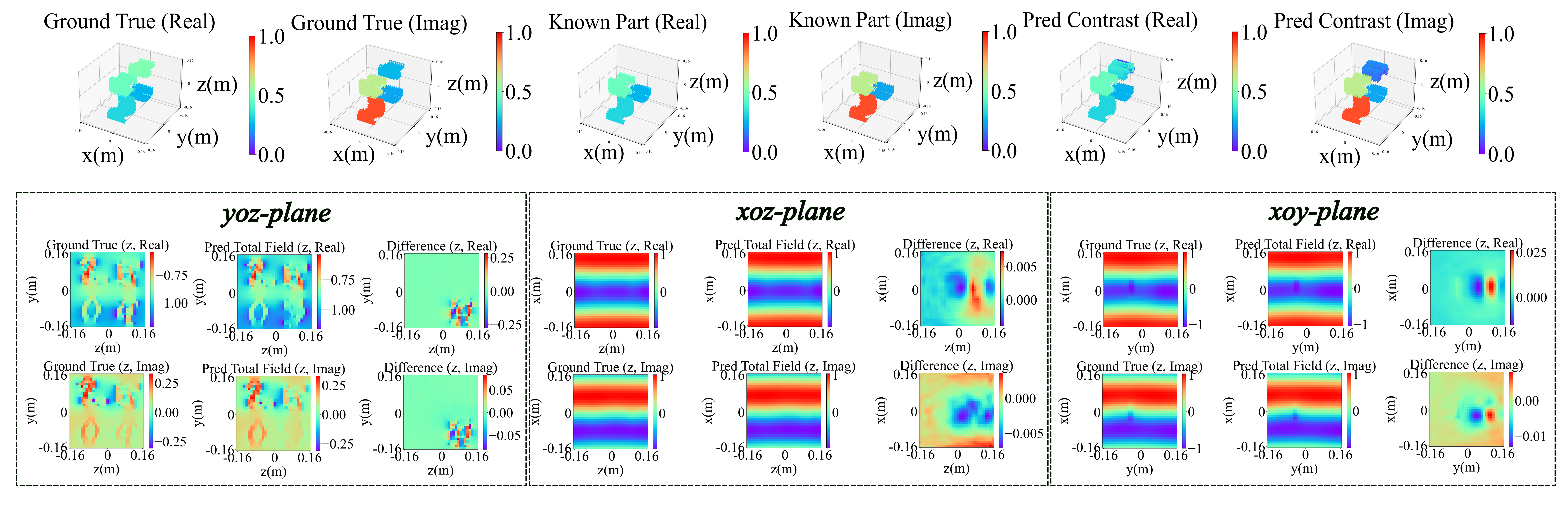}
\caption{3-D scatterers: predicted results.
$\bm{\chi}^\text{Label}$ as well as $\bm{\chi}^{p1}$ are shown in the first row.
The last two columns in the first row give the real and imaginary parts of $\bm{\chi}^\text{Pred}$.
The remaining content shows the real and imaginary part of $\mathbf{E}^\text{tot}_z$.
Specifically, the first and second columns display $\mathbf{E}^\text{tot}_z$ in the $yoz$-plane.
The forth and fifth columns display $\mathbf{E}^\text{tot}_z$ in the $xoz$-plane.
The seventh and eighth columns display $\mathbf{E}^\text{tot}_z$ in the $xoy$-plane.
The third, sixth and ninth columns present the difference between the predicted and the MoM results.
}
\label{fig:pred_result_3d_et}
\end{figure*}

\begin{figure}[!tpb]
  \centering
  \subfigure[On a circle with a radius of 3 m in the $xoz$-plane.\label{fig:Es_3d_3}]{
  \centering
  \includegraphics[width=0.48\textwidth]{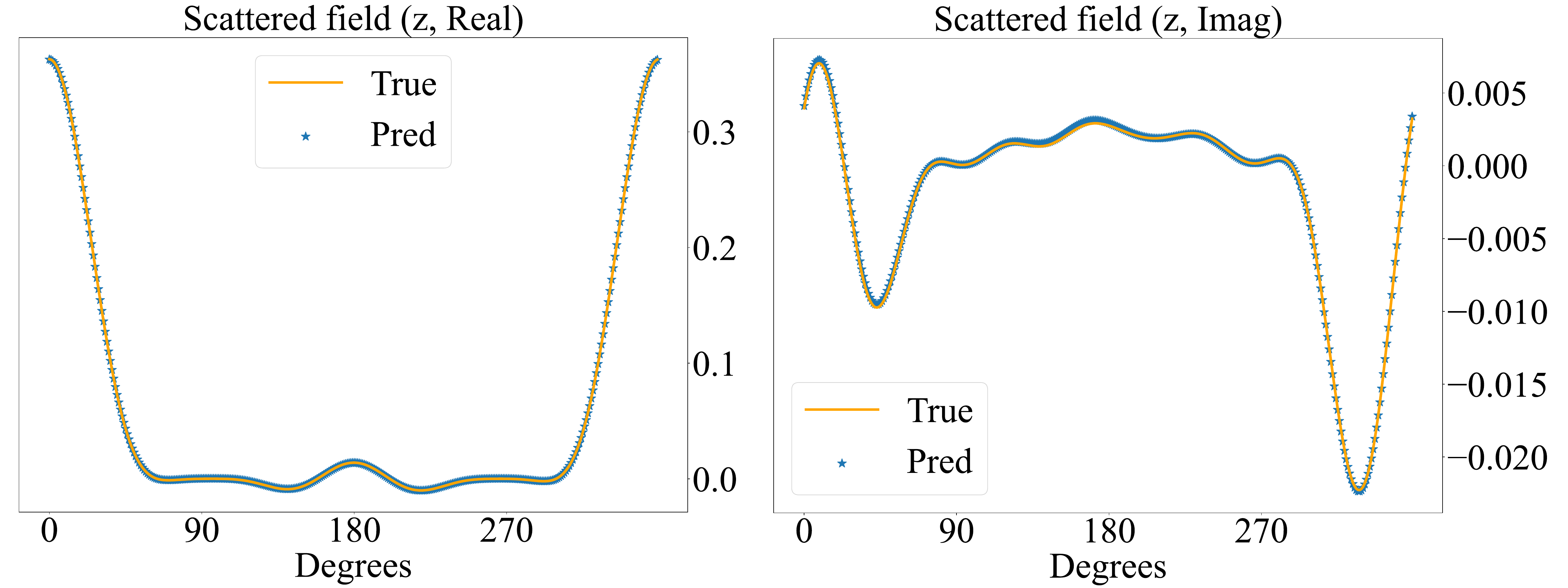}}
  \subfigure[On a circle with a radius of 10 m in the $xoz$-plane.\label{fig:Es_3d_10}]{
  \centering
  \includegraphics[width=0.48\textwidth]{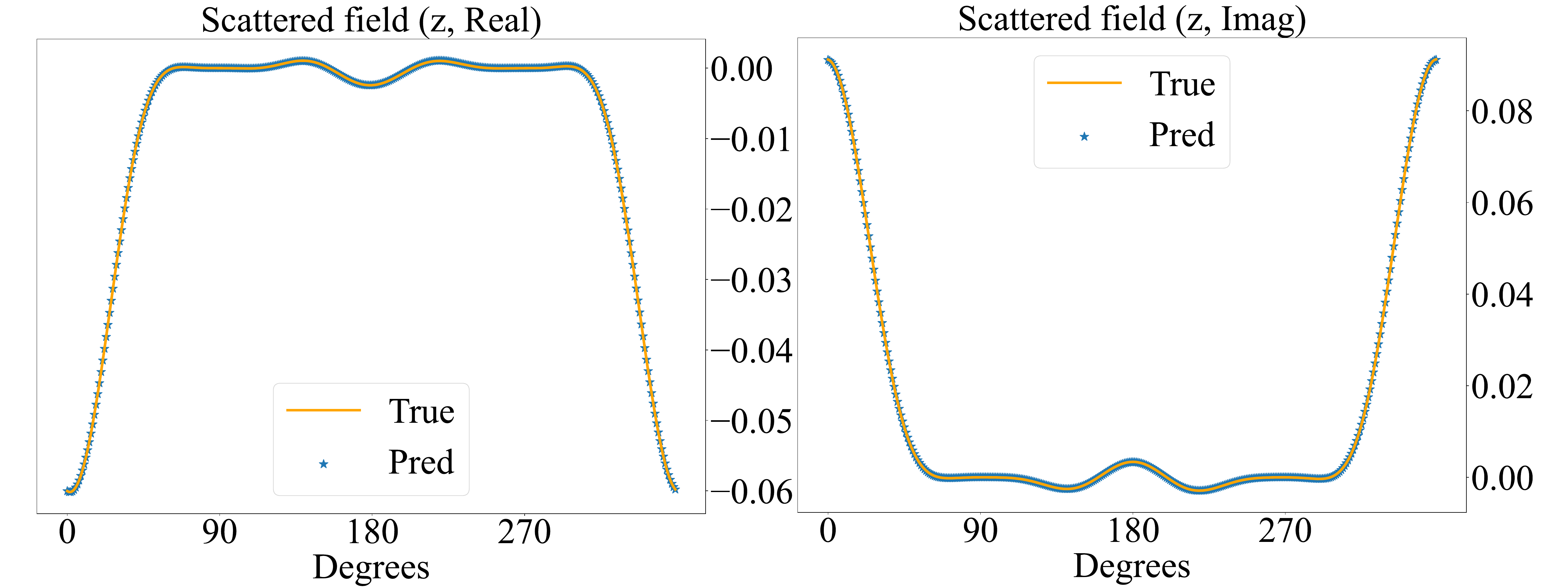}}
\caption{3-D scatterer given in Fig.~\ref{fig:pred_result_3d_et}: $\mathbf{E}^\text{sca}_z$ results. 
The first and second columns give the real and imaginary part of $\mathbf{E}^\text{sca, Pred}_z$ within [$0^\circ$, $360^\circ$], respectively.
}
\label{fig:pred_result_3d_es}
\end{figure}

\subsection{Validation}
Both datasets for 2-D and 3-D MNIST scatterers are generated by a similar manner. 
Taking the former for example, a scatterer set consisting of dielectric digit-shaped-scatterers is generated according to MNIST. 
The real and imaginary components of the contrast of each digit-shaped-scatterer are randomly selected from 0.10-1.00 and 0.00-1.00.
By randomly selecting four different digit-shaped-scatterers from the scatterer set, a sample is obtained.
One such 2-D sample is presented in Fig.~\ref{fig:esc_2d}. 
Repeating the process, a dataset having 16,000 samples is produced, among which 15,000 ones are in the training set and the rest are in the testing set.
For every sample in the dataset, three randomly chosen digit-shaped-scatterers are set as the known part $p1$ and that of the rest one is unknown part $p2$.
To compensate $\bm{\chi}^{p2}$, $\mathbf{E}^\text{sca}_\text{0}$ under a plane wave illumination is simulated by MoM with the fast Fourier transform (FFT) acceleration~\cite{10103814}. 
As a result, for each sample in the dataset, $\mathbf{E}^\text{sca}_\text{0}$ and $\bm{\chi}^{p1}$ are what we can offer the neural network as the input, along with the corresponding label $\bm{\chi}^{p2}$.

In this subsection, the working frequency is 1.0 GHz.
Since $\mathbf{E}^\text{sca}_z$ is about $10^2$ larger than its $x$- and $y$-components in magnitude (if exists), only $z$-components of the scattering and total fields are visualized in this subsection if not specified particularly.

\subsubsection{2-D Cases}\label{sec:fv2d}
The suitable choice of the input-output pair for the proposed scheme along with its performance is investigated by numerical experiments in terms of 2-D scatterers, where 2-D U-Nets~\cite{ronneberger2015u} are utilized as the backbones. 

Each 2-D sample has the computational domain $\mathcal{D}$ with a dimension of $0.64\times0.64$ m$^2$, centering at (0, 0) m. 
A $64 \times 64$ grid is employed, leading to $0.01 \times 0.01$ m$^2$ rectangular cells.
Illuminated by a transverse magnetic (TM) wave coming from the direction of 0$^\circ$ with respect to the $+x$-axis, $\mathbf{E}^\text{sca}_\text{0}$ is sampled by uniformly placing $N_s=32$ receivers on a circle with a radius of 5 m.
During the testing stage, $\mathbf{E}^\text{sca, Pred}$ is obtained at $N_r=360$ observation points that are uniformly located on a circle with a radius of 10 m.


Four input-output pairs are studied here. %
As expected, since the number of epochs is fixed, the associated training time does not vary much with the employed input-output pair, each costing about 7 hours (h).
At the same time, the prediction cost for each scatterer is around 0.006 seconds (s), not varying much either. 
It should be pointed out that it would take MoM around 0.017 s to solve Eq.~\eqref{eq:dis_j} when both $\bm{\chi}^{p1}$ and $\bm{\chi}^{p2}$ are available.
Table~\ref{tab:di_test3} gives the associated MREs.

As shown by Table~\ref{tab:di_test3}, $\bm{\chi}^{p1}$ is not fed into Network-I in the Pair I case, meanwhile $\mathbf{E}^\text{sca}_0$ is not employed in the Pair II case.  
The errors for $\bm{\chi}^\text{Pred}$ in these two cases are quite high, reaching to about 20.63\% and 15.27\%. 
In contrast, when both $\bm{\chi}^{p1}$ and $\mathbf{E}^\text{sca}_0$ are utilized, the error is decreased to 9.32\% as shown by the Pair III case.
It should be noted that, in the Pair I case, $\bm{\chi}^{p1}$ is employed in Network-II by fixing the corresponding values in $\bm{\chi}$. 
Since both $\bm{\chi}^{p1}$ and $\mathbf{E}^\text{sca}_0$ are utilized, the cases of Pair III and Pair IV can reach lower errors than that of Pair I and Pair II. 
Different from the Pair IV case, $\mathbf{E}^\text{inc}$ is replaced by $\mathbf{E}^{p1}$ in the Pair III case.
The fact that the latter case exhibits the smaller prediction errors than the former one reveals that the explicit employment of $\mathbf{E}^{p1}$ as shown in Eq.~\eqref{eq:lcon} can improve the performance of the retrieval, consistent with our analysis in Section~\ref{sec:Formulation}.

To further validate our statement in Section~\ref{sec:Formulation} that scattering data under a single or a very few number of incident waves are insufficient for traditional solvers to retrieve $\bm{\chi}^{p2}$ accurately, the SOM results are compared with that of the proposed DL-based scheme when Pair III is employed in Fig.~\ref{fig:pred_result}. 
The SOM employed here is modified so that $\bm{\chi}^{p1}$ is utilized to fix the contrast values within $p1$ during the solution process.
That is to say, only $\bm{\chi}^{p2}$ is iteratively retrieved in the SOM.
The MRE of $\bm{\chi}^\text{Pred}$ obtained by the SOM is 68.27\%, demonstrating the SOM incapable of retrieving $\bm{\chi}^{p2}$ accurately. 

\begin{table}[!t]
  \begin{center}
    \caption{REs for computations presented in Figs.~\ref{fig:pred_result_3d_et} and~\ref{fig:pred_result_3d_es}.
    }
    \label{tab:di_test333}
    \setlength{\tabcolsep}{8mm}
    \renewcommand\arraystretch{1.3}
    \begin{tabular}{|cc|}
      \hline
      \multicolumn{1}{|c|}{$\bm{\chi}^\text{Pred}$ (\%)} & 5.24   \\ \hline
      \multicolumn{1}{|c|}{$\mathbf{E}^\text{tot, Pred}$ (\%)}    & 2.70   \\ \hline
      \multicolumn{1}{|c|}{$\mathbf{E}^\text{sca, Pred}$ (3 m, \%)}    & 0.18   \\ \hline
      \multicolumn{1}{|c|}{$\mathbf{E}^\text{sca, Pred}$ (10 m, \%)}    & 0.17    \\ \hline
      \end{tabular}
  \end{center}
\end{table}

\begin{figure}[!tpb]
  \centering
  \subfigure[The ground true.]{
  \includegraphics[width=0.35\textwidth]{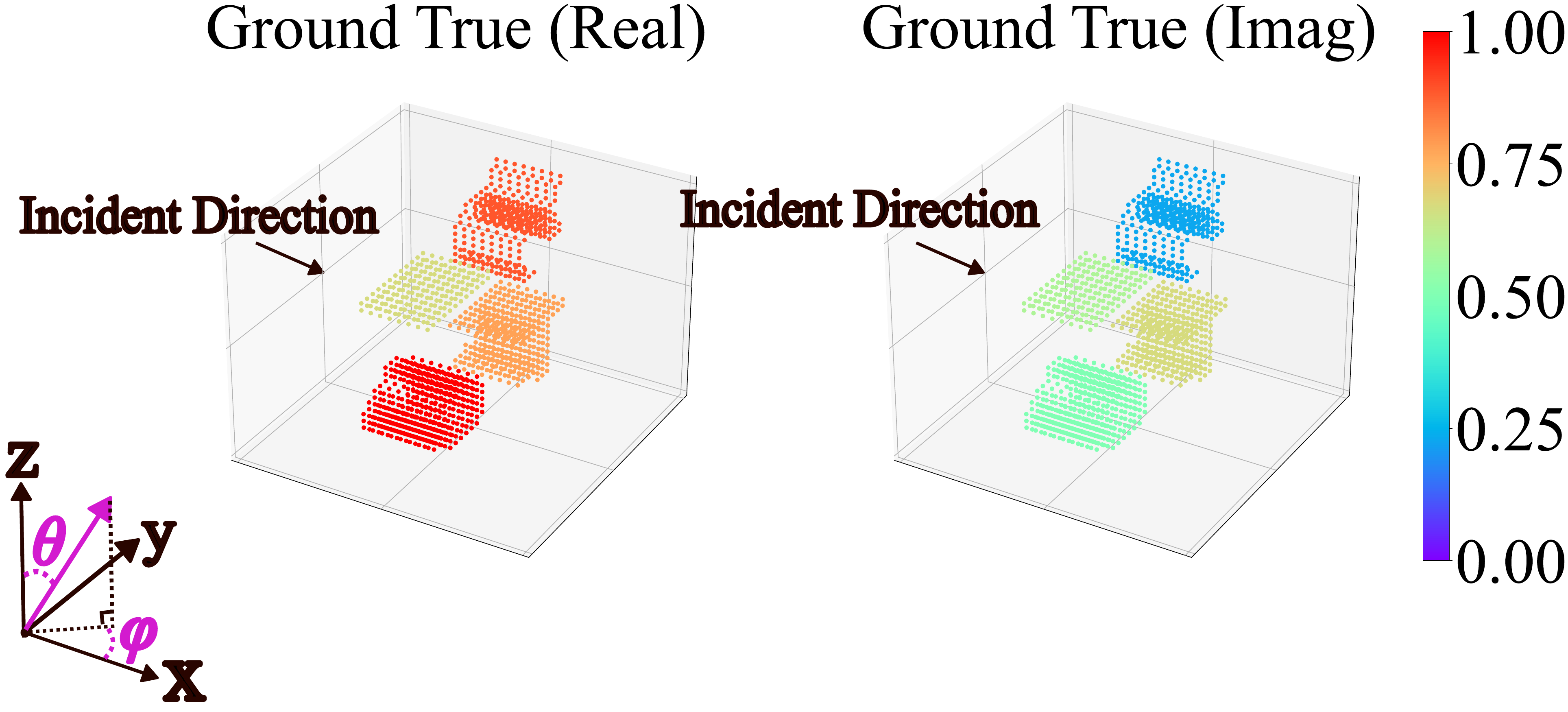}
  \label{fig:xi_true}
  }
  \subfigure[Case-I.]{
  \includegraphics[width=0.35\textwidth]{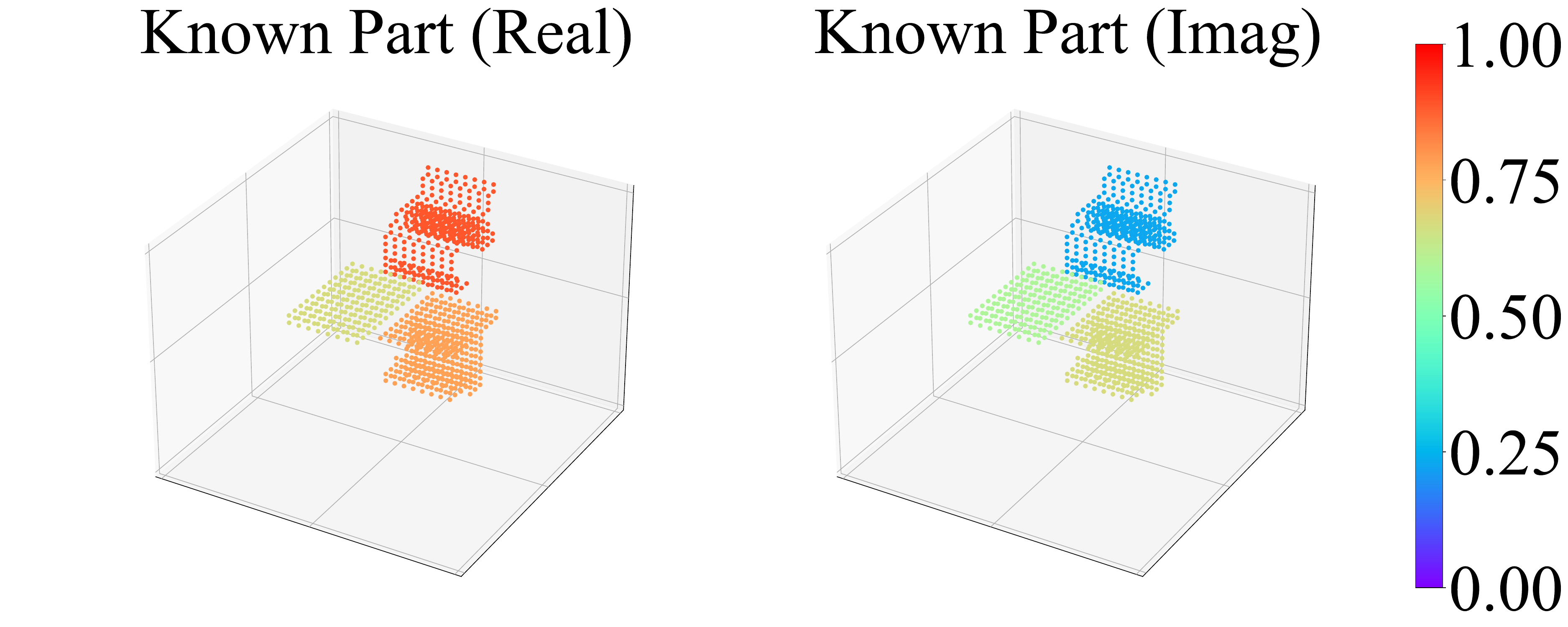}
  \label{fig:xi_case1}
  }
  \subfigure[Case-II.]{
  \includegraphics[width=0.35\textwidth]{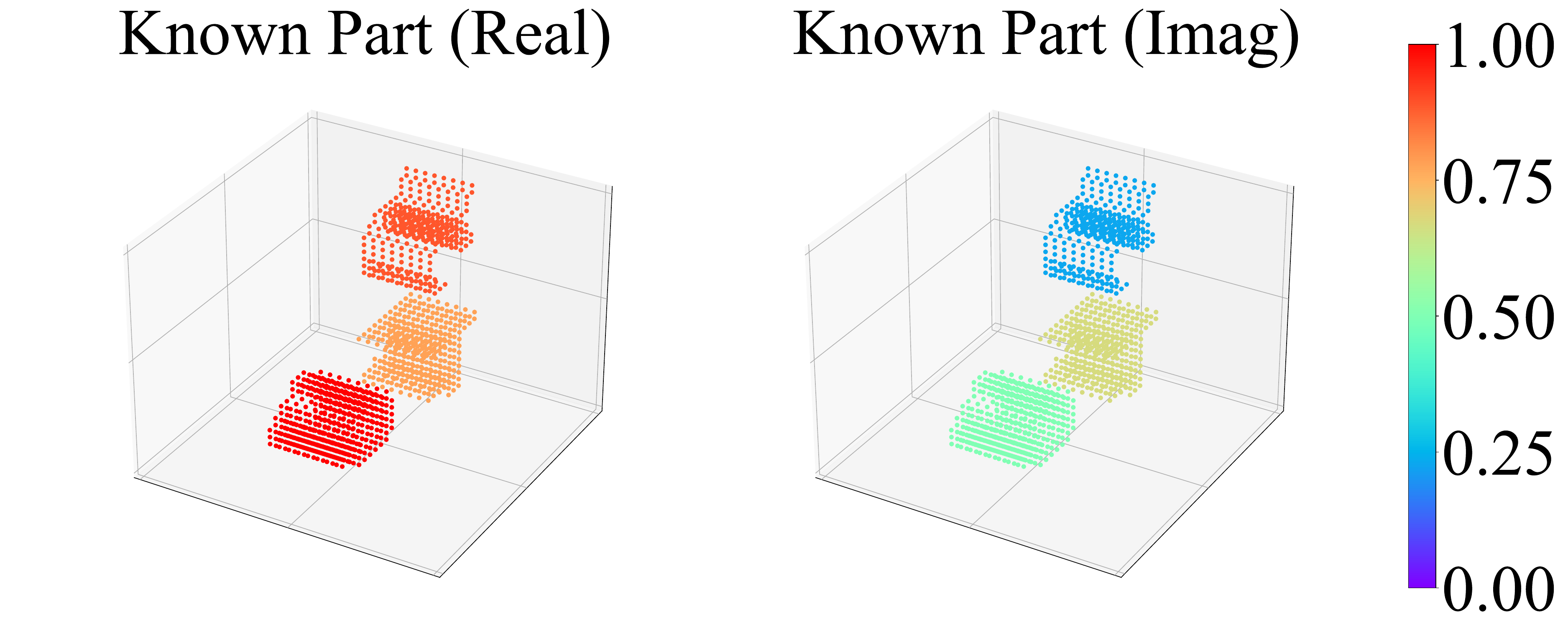}
  \label{fig:xi_case2}
  }
\caption{Generalization ability in terms of the incident direction: MNIST objects under investigation.
}
\label{fig:imdia}
\end{figure}

\begin{figure}[!t]
  \centering
  \subfigure[Case-I: REs of $\mathbf{{E}}^\text{sca, Pred}$ (3 m). ]{
  \includegraphics[width=0.22\textwidth]{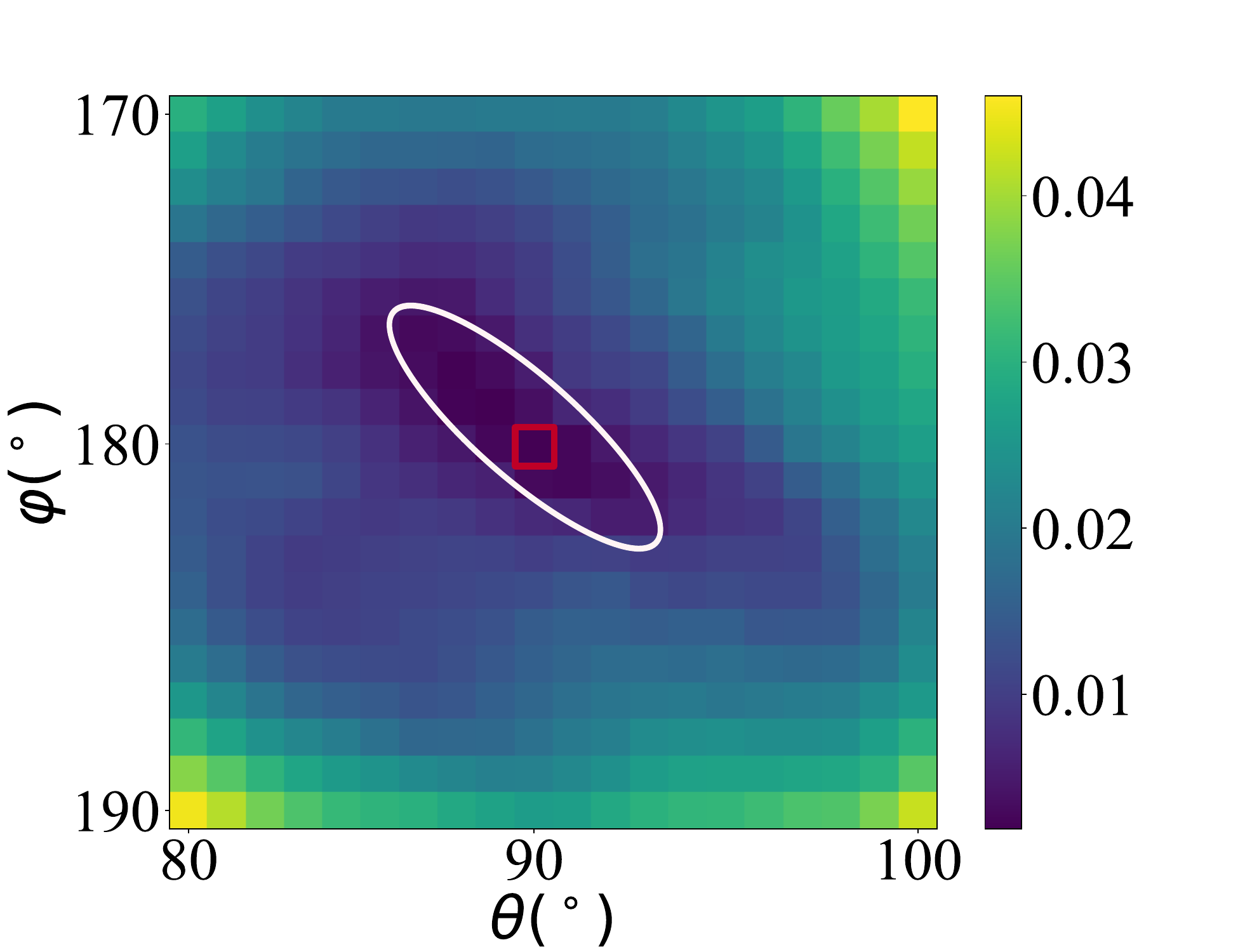}
  \label{fig:re_esca_3}
  }
  \subfigure[Case-I: REs of $\mathbf{{E}}^\text{sca, Pred}$ (10 m).]{
  \includegraphics[width=0.22\textwidth]{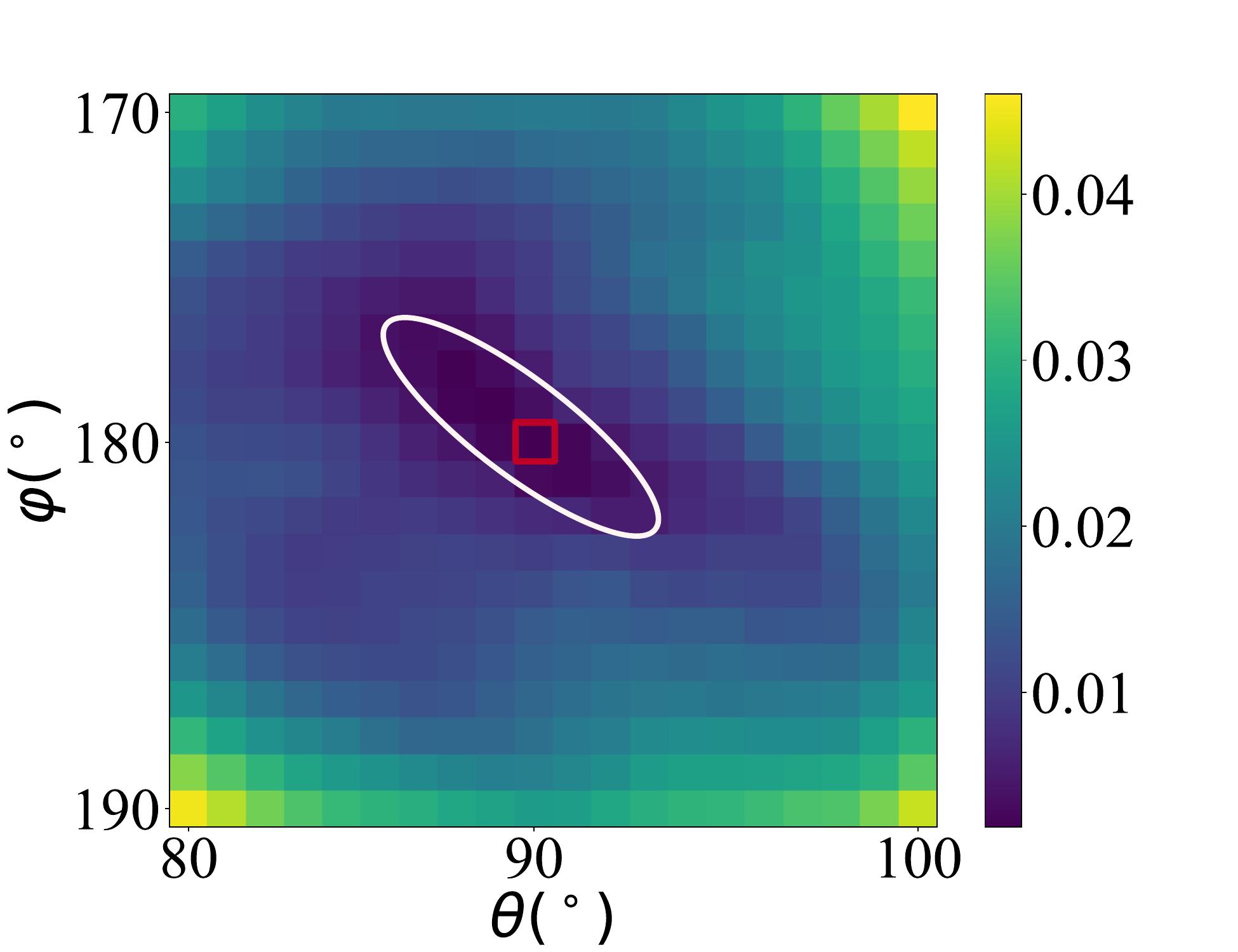}
  \label{fig:re_esca_10}
  }
  \subfigure[Case-II: REs of $\mathbf{{E}}^\text{sca, Pred}$ (3 m). ]{
  \includegraphics[width=0.22\textwidth]{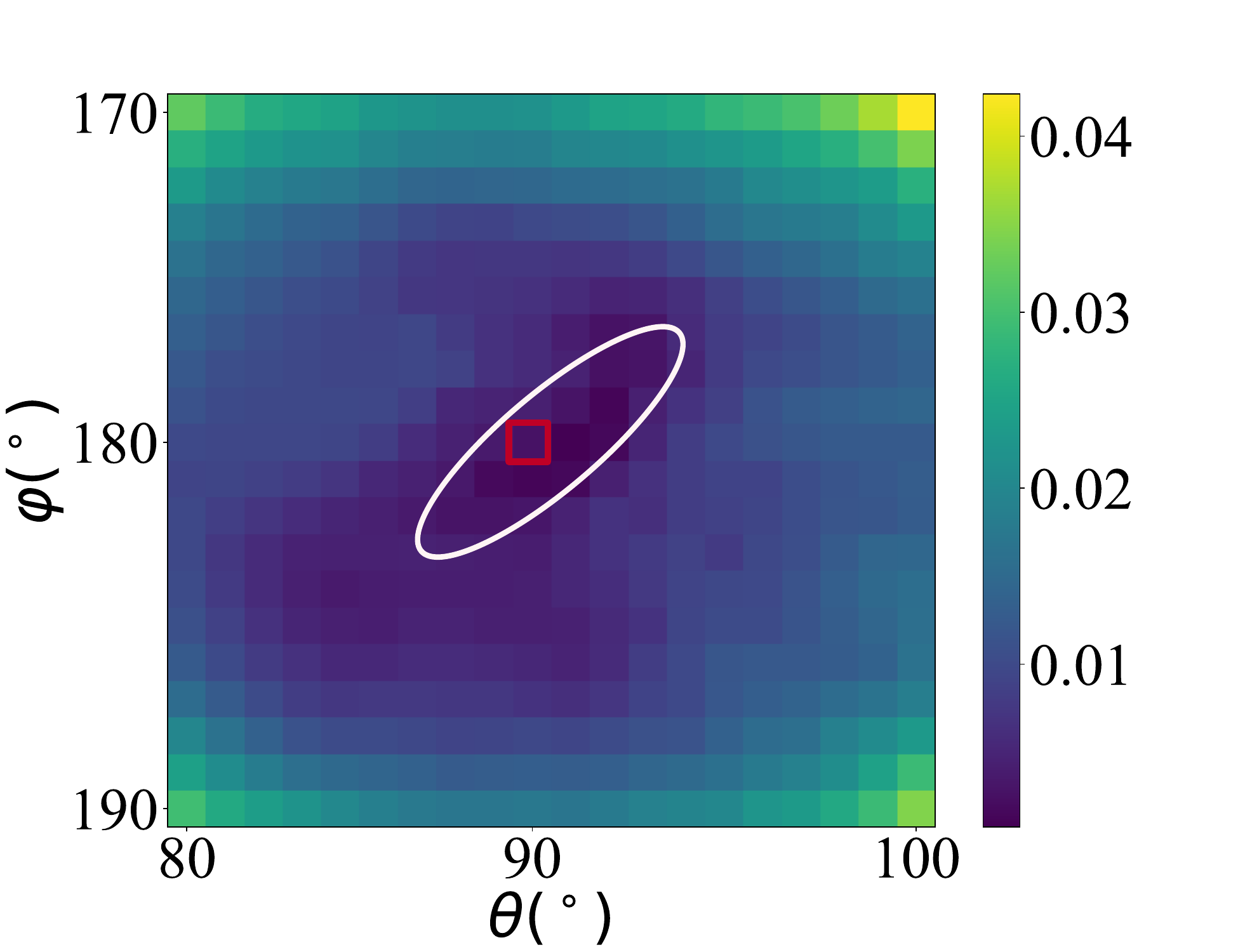}
  \label{fig:re_esca_3_case2}
  }
  \subfigure[Case-II: REs of $\mathbf{{E}}^\text{sca, Pred}$ (10 m).]{
  \includegraphics[width=0.22\textwidth]{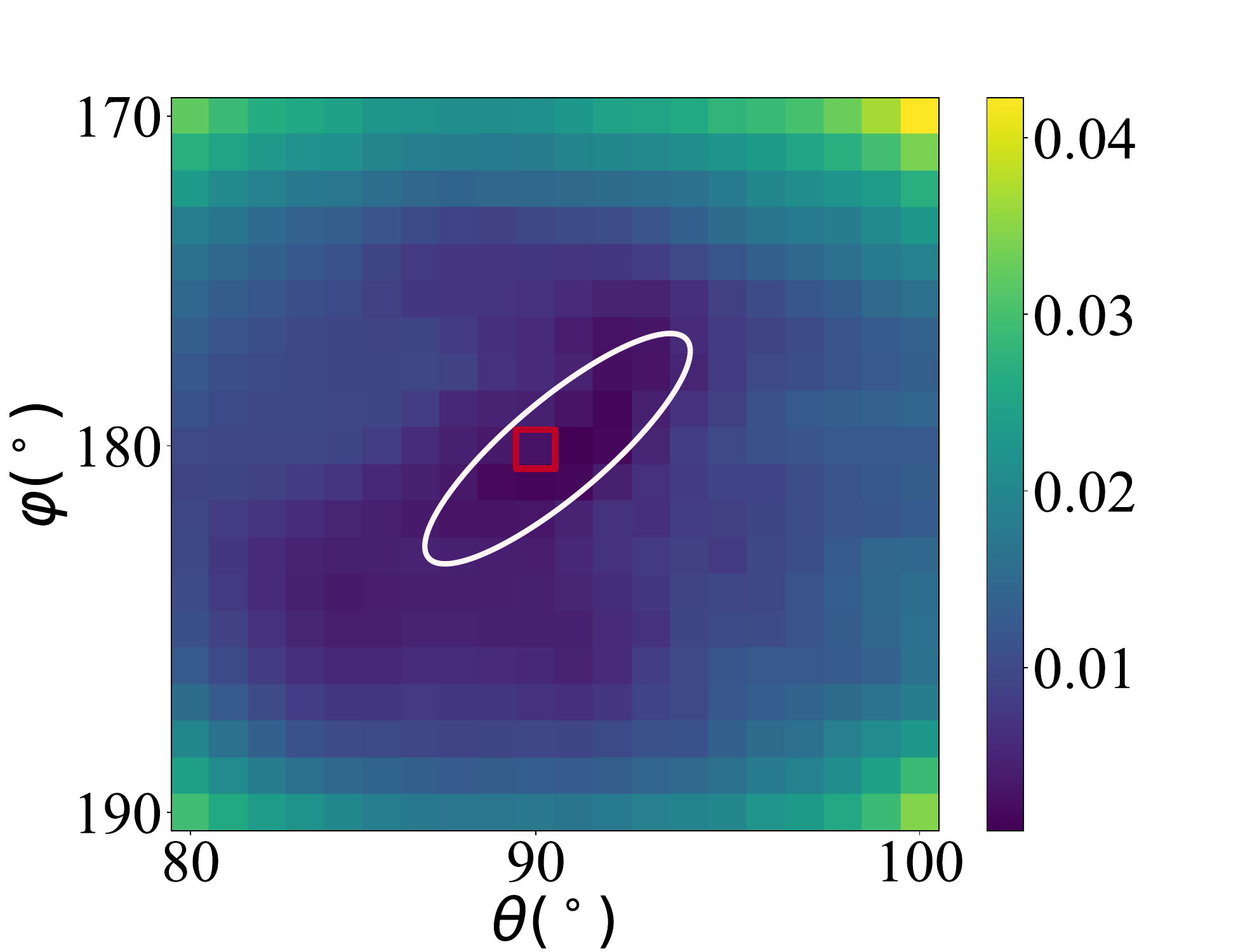}
  \label{fig:re_esca_10_case2}
  }
\caption{Generalization ability in terms of the incident direction: REs of $\mathbf{{E}}^\text{sca, Pred}$'s for the object shown in Fig.~\ref{fig:imdia} under different incident directions.
}
\label{fig:re_e}
\end{figure}

\begin{figure}[tpb]
  \centering
  \includegraphics[width=0.48\textwidth]{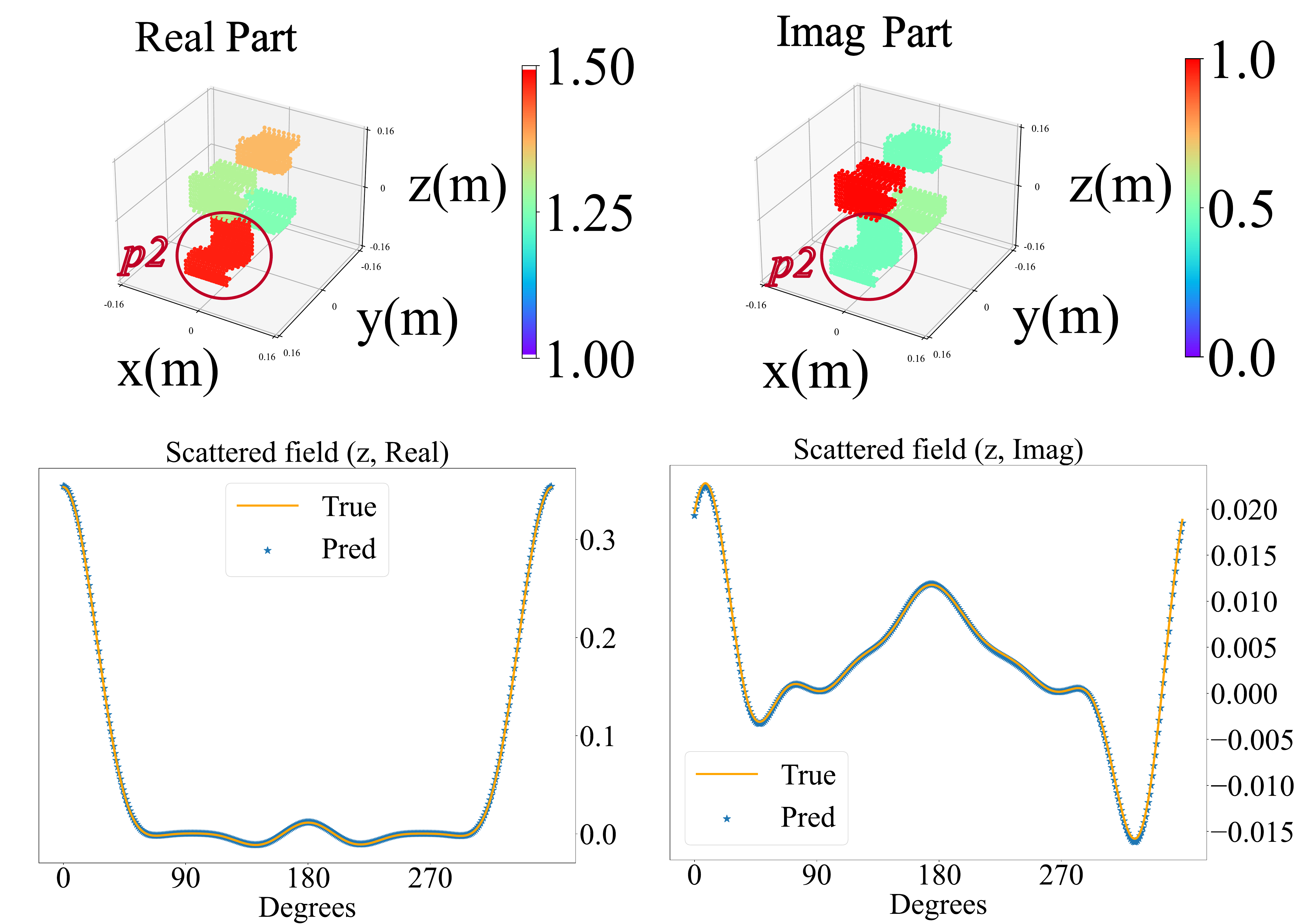}
\caption{Generalization ability in terms of the contrast: $\mathbf{{E}}^\text{sca}_z$ results for a randomly selected scatterer. 
The testing points are on a circle with a radius of 3 m in the $xoz$-plane. 
}
\label{fig:pred_result_3d_es_1_1p5}
\end{figure}

\subsubsection{3-D Cases}
\label{sec:fv3d}

In this study, each sample has the computational domain $\mathcal{D}$ with a dimension of $0.32\times0.32\times0.32$ m$^3$, centering at (0, 0, 0) m. 
Voxels with size $0.01 \times 0.01 \times 0.01$ m$^3$ are utilized to discretize the domain $\mathcal{D}$.
Such a 3-D sample is provided in Fig.~\ref{fig:esc}.
By setting the incident direction being $(\theta=90^\circ, \phi=180^\circ)$, the input $\mathbf{E}^\text{sca}_\text{0}$ is sampled by receivers that are uniformly located on circles in the $xoy$-, $xoz$- and $yoz$-plane, respectively, each with a radius of 5 m. 
There are 64 receivers on each circle, resulting in $N_s=64\times 3=192$.
The incident field is a $z$-polarized plane wave propagating towards +$x$-axis.
%
Pair III is employed during the study. 
The training costs around 91 h.
A single prediction process can finish in around 0.025 s, compared to around 1.5 s for MoM to solve Eq.~\eqref{eq:dis_j} when both $\bm{\chi}^{p1}$ and $\bm{\chi}^{p2}$ are available.
In the testing stage, $\mathbf{E}^\text{sca, Pred}$ is obtained on two circles with radii of 3 m and 10 m in the $xoz$-plane, each having $N_r=360$ uniformly distributed observation points. 
MRE of ${\bm{\chi}^{{\text{Pred}}}}$ reaches 6.41\% and that of $\mathbf{E}^\text{tot, Pred}$ reaches 3.20\%. 
At the same time, MRE of $\mathbf{E}^\text{sca, Pred}$ on the circle of 3m radius reaches 0.15\% and that of 10m radius is 0.14\%.

Figure~\ref{fig:pred_result_3d_et} presents $\mathbf{E}^\text{tot, Pred}_z$
 for the testing sample randomly selected from the testing set while Fig.~\ref{fig:pred_result_3d_es} gives $\mathbf{E}^\text{sca, Pred}_z$. 
The corresponding testing errors are listed in Table~\ref{tab:di_test333}.
Again, it can be seen that the proposed DL-based scheme can efficiently retrieve $\bm{\chi}^{p2}$ and then predict the scattering field accurately. 

\begin{figure*}[!tpb]
  \centering
  \includegraphics[width=0.98\textwidth]{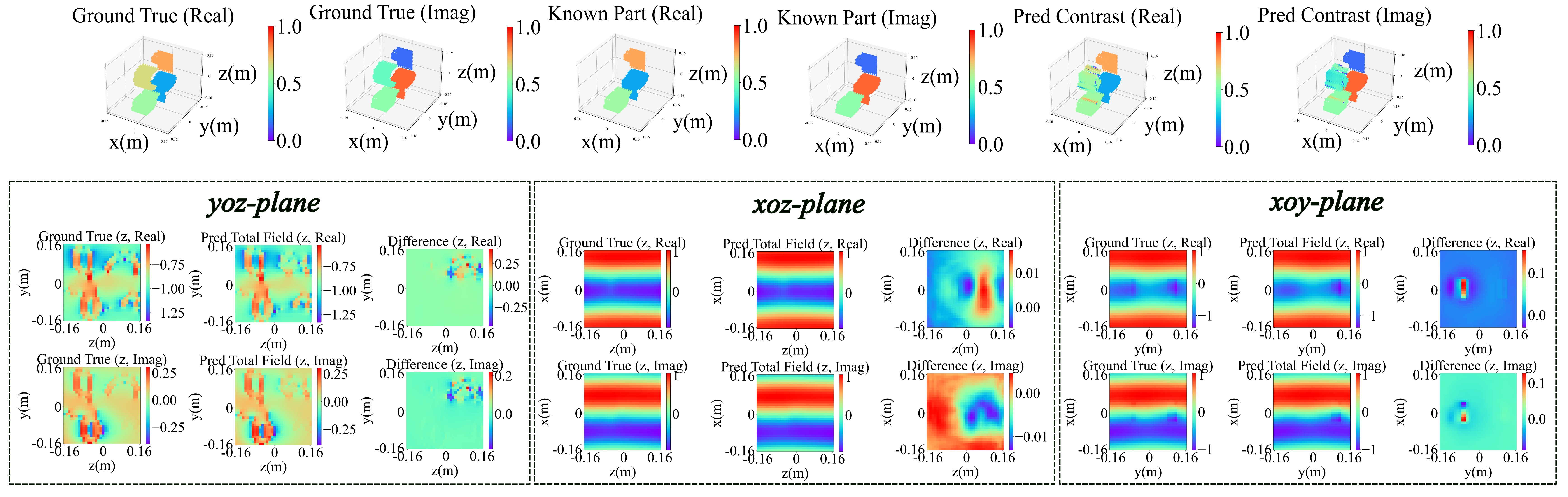}
\caption{Generalization ability in terms of the geometry shape: the predicted results for a scatterer generated according to EMNIST.
}
\label{fig:pred_result_ds}
\end{figure*}

\begin{figure}[!tpb]
  \centering
  \subfigure[The scattering field on a circle with a radius of 3 m in the $xoz$-plane.\label{fig:Esemnist_3d_3}]{
  \centering
  \includegraphics[width=0.48\textwidth]{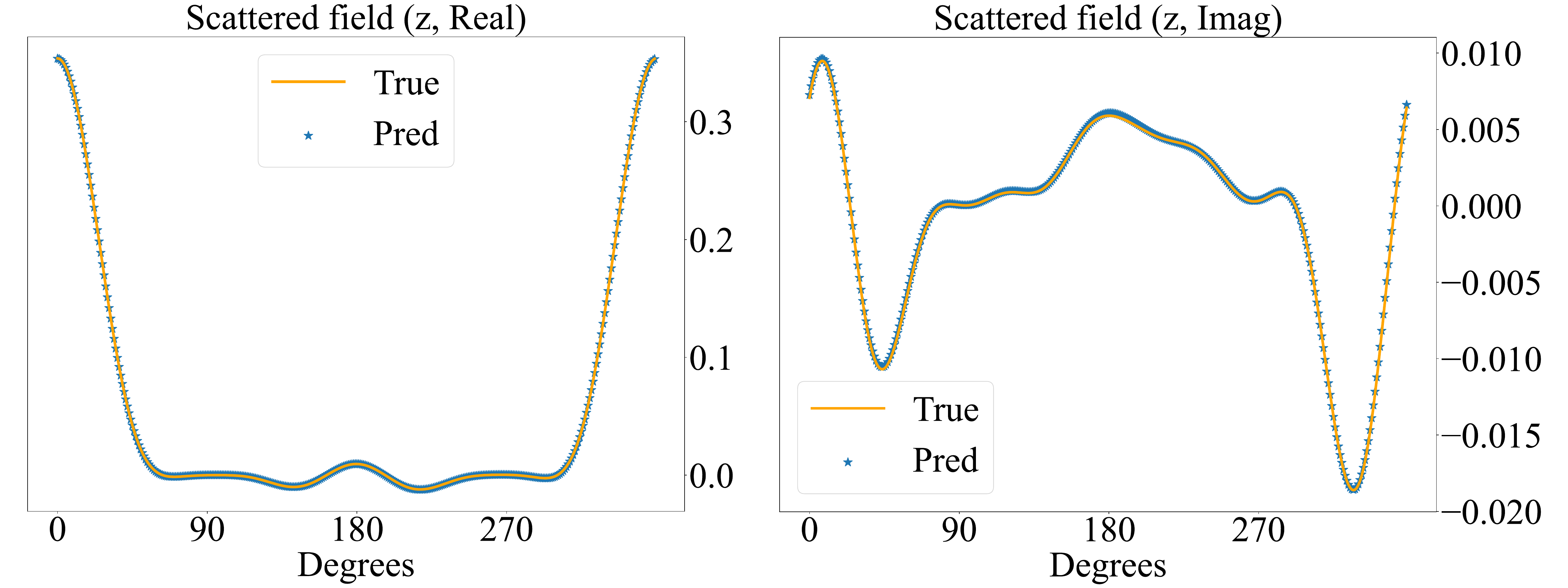}}
  \subfigure[The scattering field on a circle with a radius of 10 m in the $xoz$-plane.\label{fig:Esemnist_3d_10}]{
  \centering
  \includegraphics[width=0.48\textwidth]{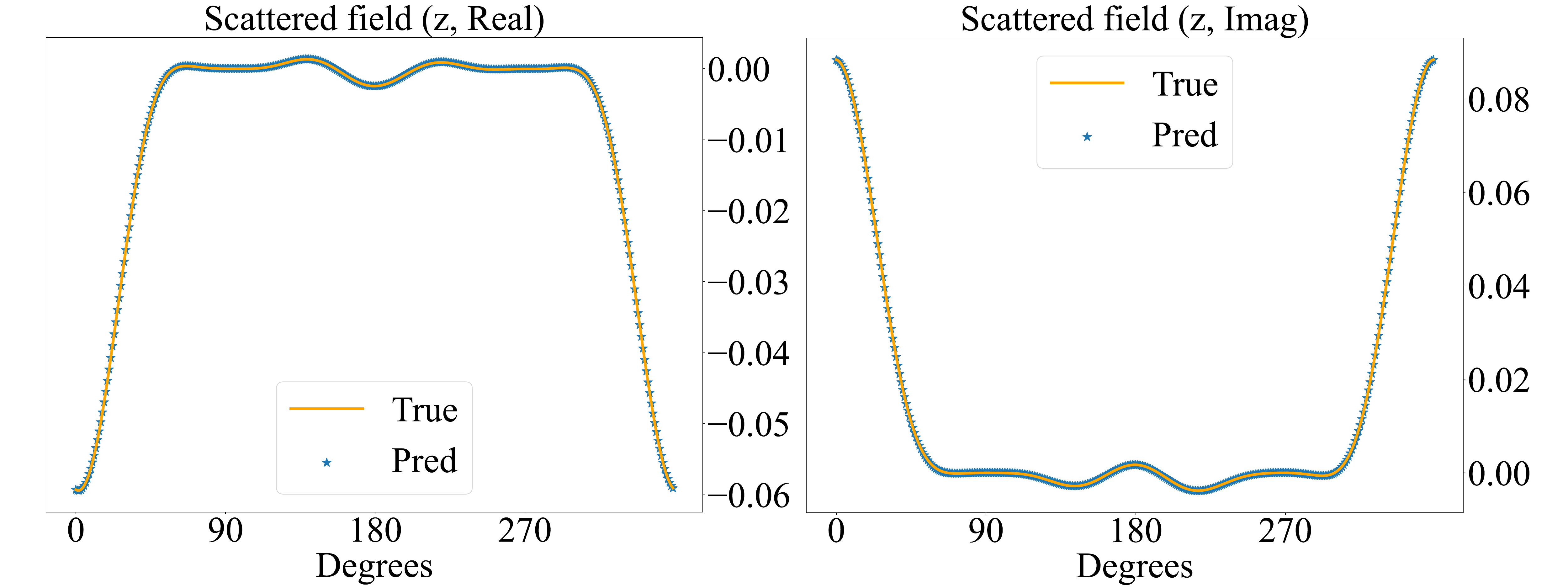}}
\caption{Generalization ability in terms of the geometry shape: the $\mathbf{E}^\text{sca}_z$ results for the EMNIST scatterer presented in Fig.~\ref{fig:pred_result_ds}. 
}
\label{fig:pred_result_3d_es_emnist}
\end{figure}

\begin{table}[!tbp]
  \begin{center}
    \caption{REs for computations presented in Figs.~\ref{fig:pred_result_ds} and~\ref{fig:pred_result_3d_es_emnist}.
    }
    \label{tab:di_test31}
    \setlength{\tabcolsep}{8mm}
    \renewcommand\arraystretch{1.3}
    \begin{tabular}{|cc|}
      \hline
      \multicolumn{1}{|c|}{$\bm{\chi}^\text{Pred}$ (\%) }  &  8.51  \\ \hline
      \multicolumn{1}{|c|}{$\mathbf{E}^\text{tot, Pred}$ (\%)}    & 3.99   \\ \hline
      \multicolumn{1}{|c|}{$\mathbf{E}^\text{sca, Pred}$ (3 m, \%)}    & 0.35   \\ \hline
      \multicolumn{1}{|c|}{$\mathbf{E}^\text{sca, Pred}$ (10 m, \%)}    & 0.35    \\ \hline
      \end{tabular}
  \end{center}
\end{table}

\subsection{Generalization}
\label{sec:discussion}

The generalization ability is studied in terms of three types of parameters, where the network for 3-D cases trained in Section~\ref{sec:fv3d} is utilized.

\subsubsection{Incident Direction}
\label{sec:sdia}
3-D scatterers discussed in subsection~\ref{sec:fv3d} are employed here. 
Since one limitation of the current implementation here is that the network can only accept $\mathbf{E}^\text{sca}_\text{0}$ with respect to a single incident wave, the incident directions in testing are restricted close to that used in generating $\mathbf{E}^\text{sca}_\text{0}$.
In particular, the variations of $\theta$ and $\varphi$ of the incident wave are limited within $\pm 10^\circ$ relative to their counterparts in the training set.
Two scenarios are discussed, where $p2$ is positioned at different places, as shown in Fig.~\ref{fig:imdia}. 
The REs of $\mathbf{E}^\text{sca, Pred}$ vary with incident $\theta$ and $\varphi$ are presented in Fig.~\ref{fig:re_e}. %
Here, two set of scattering fields are evaluated for each computation, on circles with radii of 3 m and 10 m in the $xoz$-plane, respectively. 
As shown in Fig.~\ref{fig:re_e}, the proposed scheme performs considerably well for all computations.
The generalization ability of the proposed scheme is quite good.
In Fig.~\ref{fig:re_e}, each subfigure has a red-colored square to indicate the incident direction used in the training set, namely, $(90^\circ, 180^\circ)$.
At the same time, there is a white-colored ellipse in each subfigure exhibiting the distribution of the computations with relatively small prediction errors. 
It is expected that, when the incident comes from directions where $p2$ is not occluded, the prediction error would become small compared to the other scenarios. 
A clear trend consistent with such expectation can be found, as shown by the white-colored ellipses in Fig.~\ref{fig:re_e}.

\subsubsection{Contrast}
Testing is carried out on 1,000 3-D MNIST scatterers with an out-of-range contrast.  
In particular, the real parts of the testing scatterers here are increased to [1.00, 1.50] while the imaginary parts remain unchanged, i.e., [0.00, 1.00]. 
Similarly, $\mathbf{E}^\text{sca, Pred}$ is sampled on two circles with radii of 3 m and 10 m in the $xoz$-plane, each having $N_r=360$ uniformly distributed observation points. 
MRE of ${\bm{\chi}^{{\text{Pred}}}}$ reaches 10.19\% and that of $\mathbf{E}^\text{tot, Pred}$ reaches 4.33\%. 
At the same time, MREs of $\mathbf{E}^\text{sca, Pred}$ on the circle of 3m and 10m radius both reach 0.27\%.
Figure~\ref{fig:pred_result_3d_es_1_1p5} gives the testing result for a randomly selected scatterer.
It is shown that the testing accuracy is still acceptable, although it decreases slightly with the increase of the contrast.

\begin{figure}[t]
  \centering
  \includegraphics[width=0.48\textwidth]{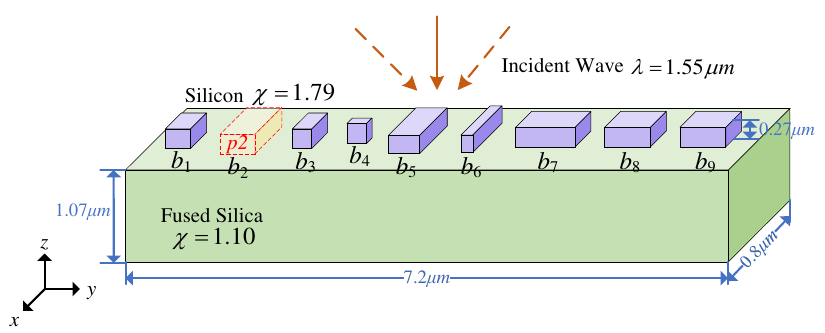}
\caption{A metasurface with an incomplete profile. }
\label{fig:metamaterial}
\end{figure}

\begin{figure*}[!t]
  \centering
  \subfigure[Top view of the testing metasurface.]{
    \includegraphics[width=0.3\textwidth]{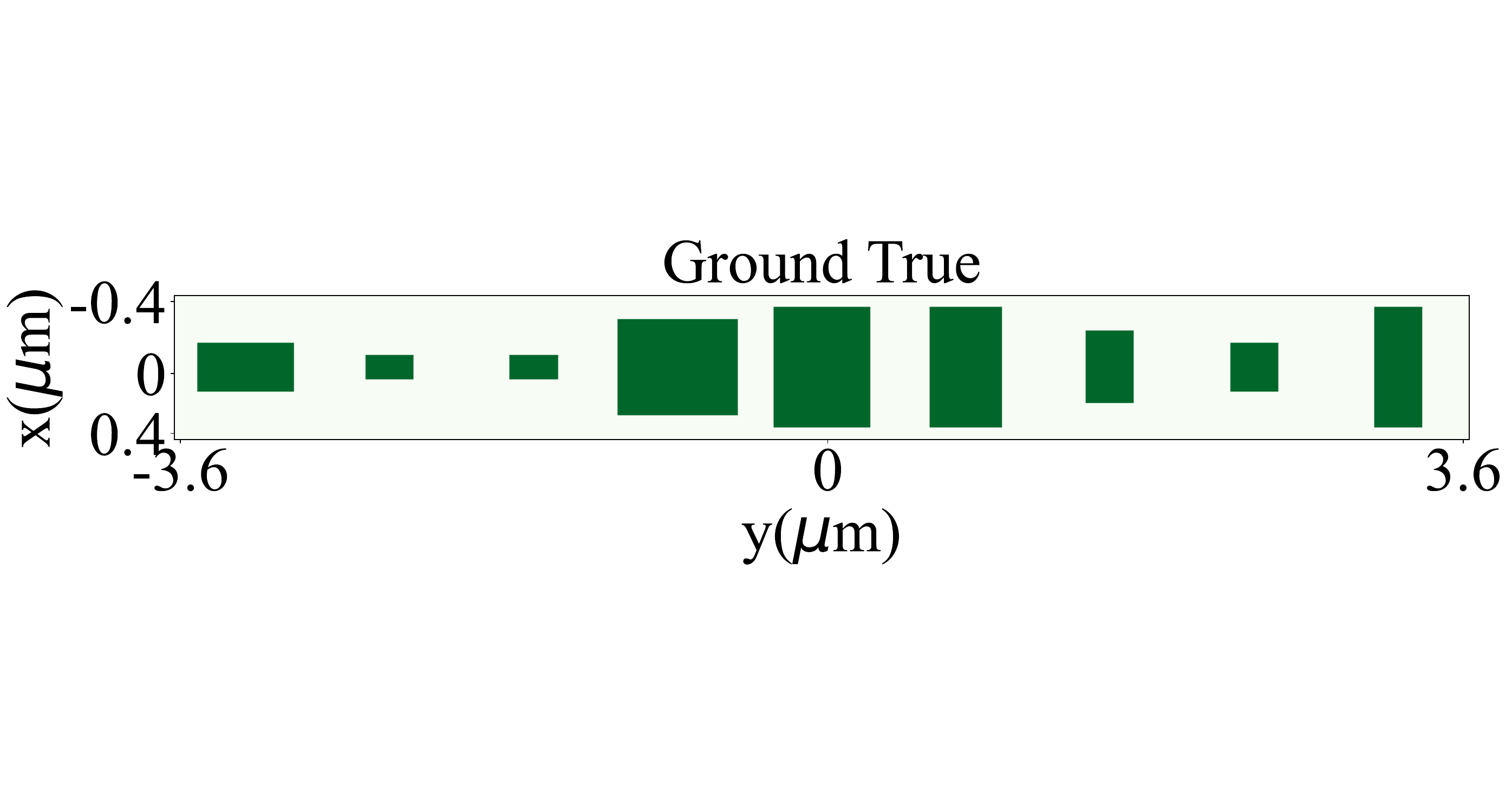}
    \label{fig:testsample}
  }
  \subfigure[Case-I]{
  \includegraphics[width=0.35\textwidth]{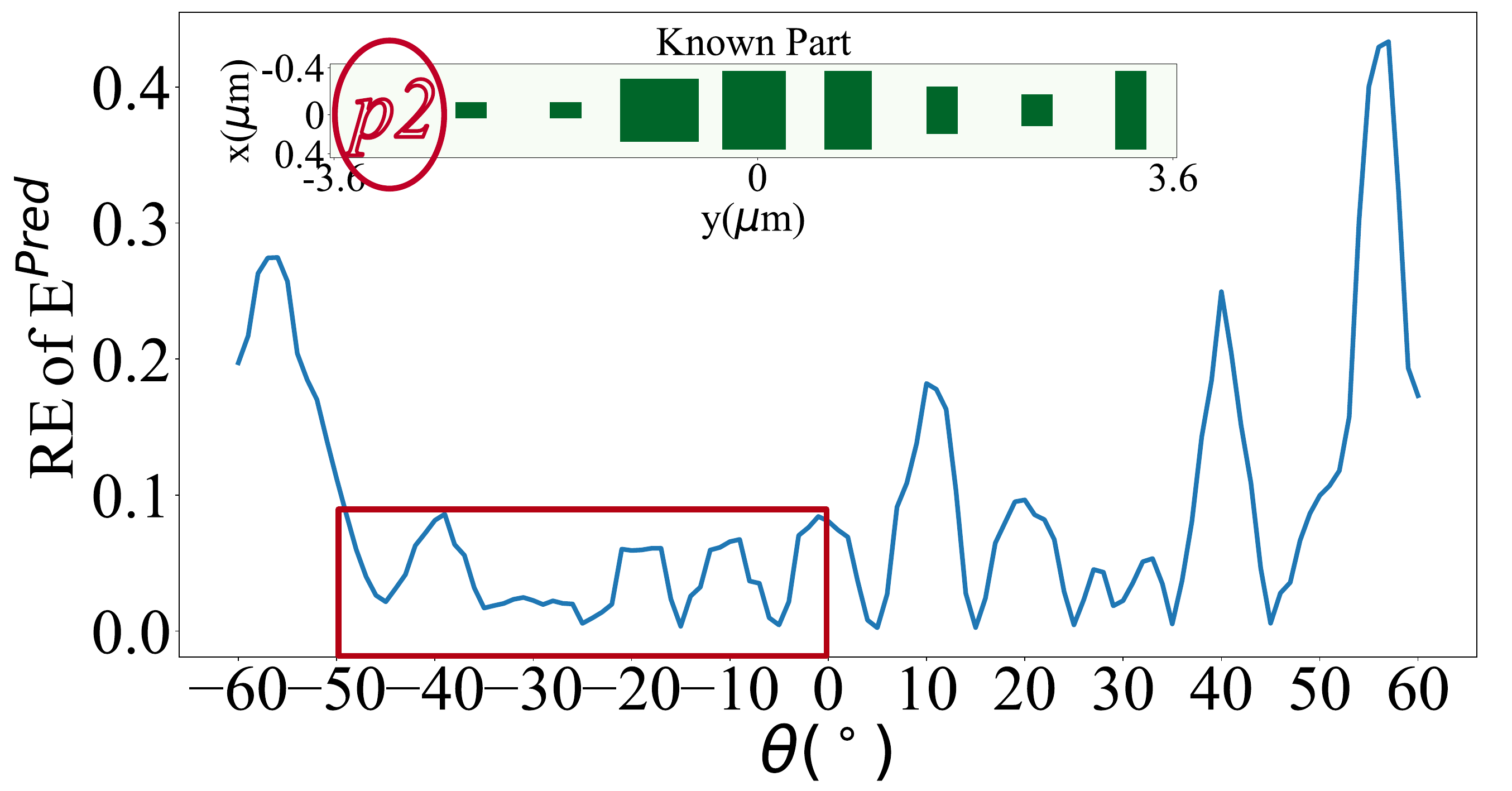}
  \label{fig:rel_error_es_left}
  }
  \subfigure[Case-II]{
  \includegraphics[width=0.35\textwidth]{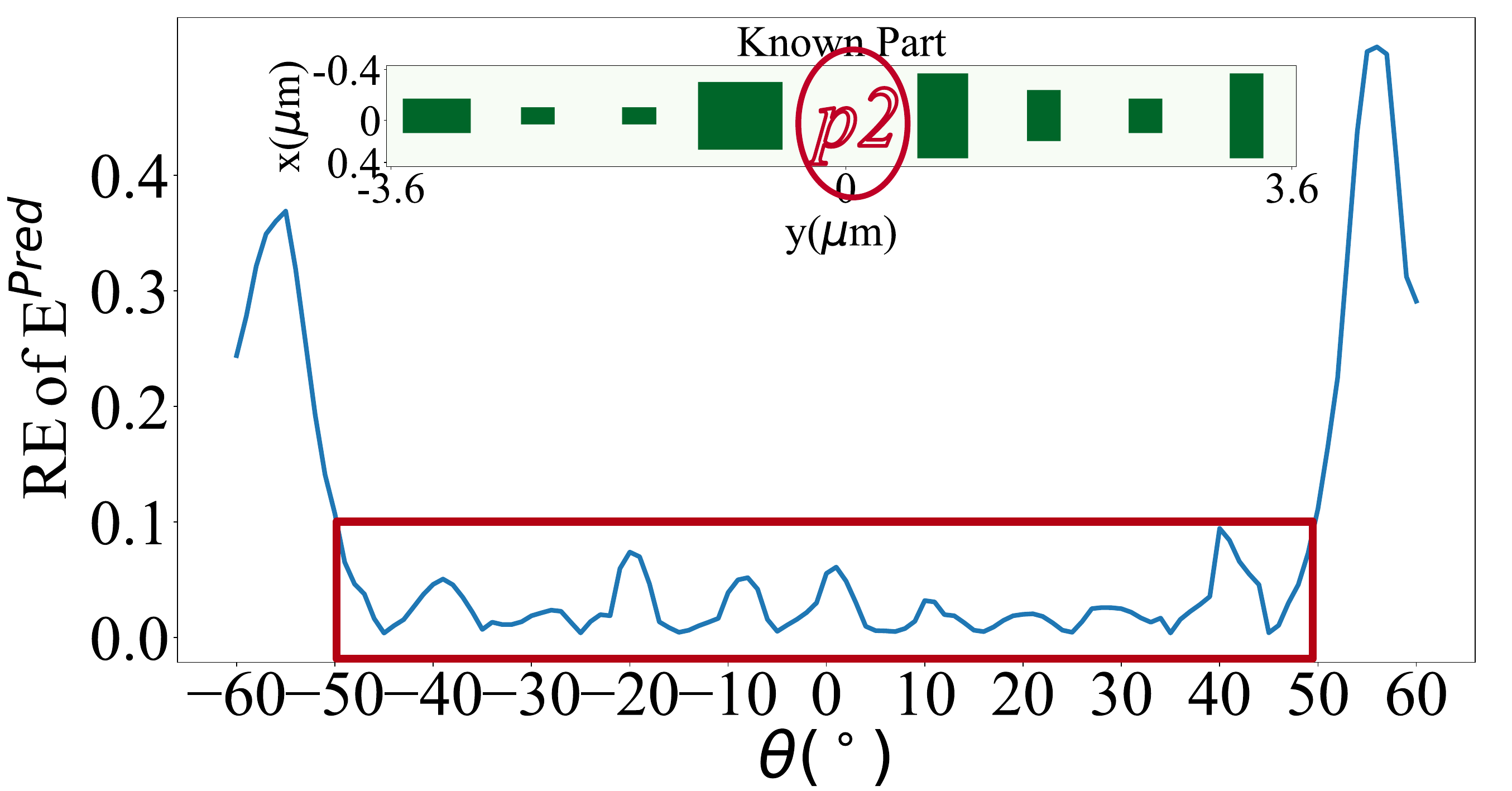}
  \label{fig:rel_error_es_center}
  }
  \subfigure[Case-III]{
  \includegraphics[width=0.35\textwidth]{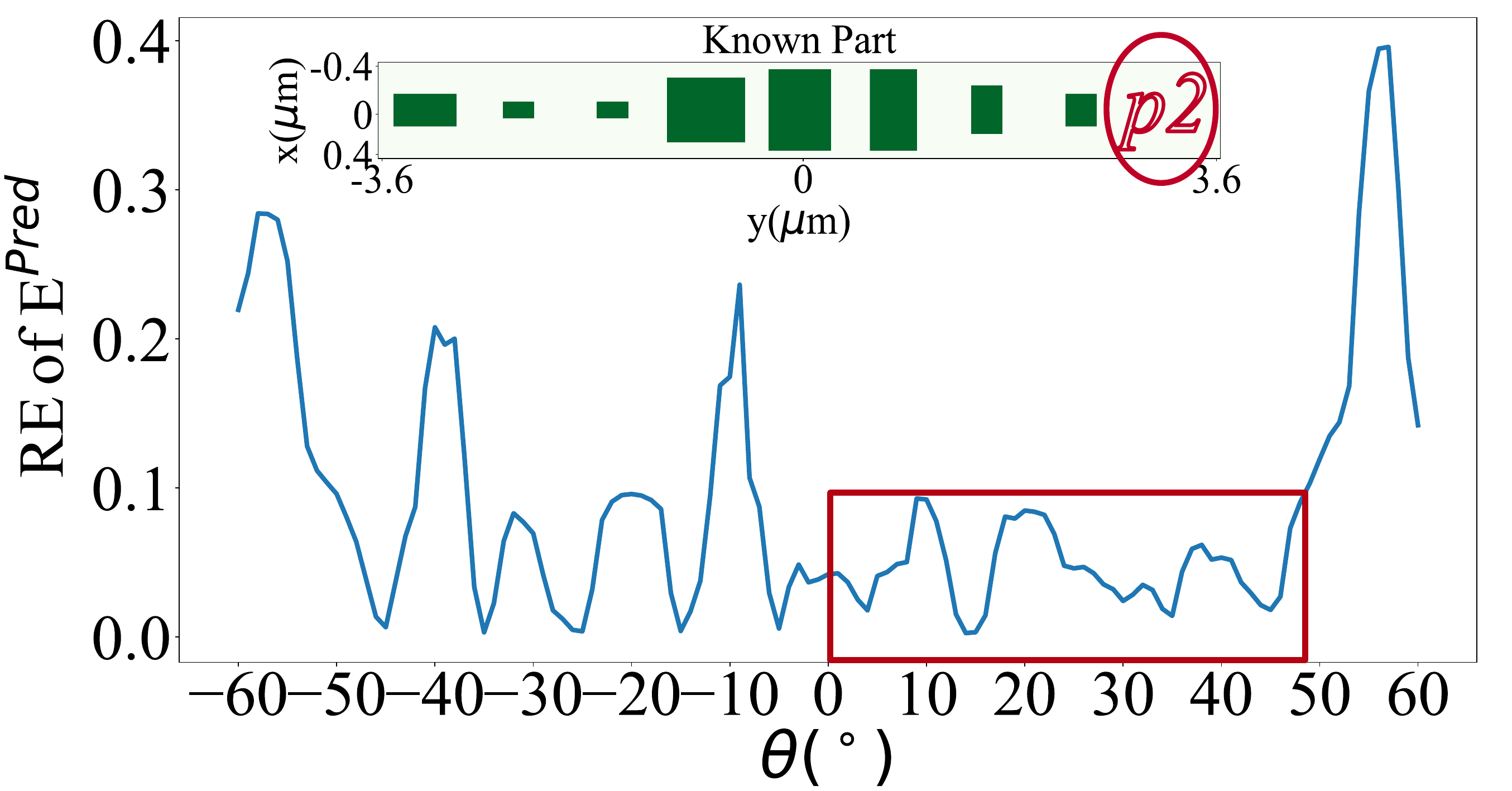}
  \label{fig:rel_error_es_right}
  }
\caption{Metasurface computations: REs vary with $\theta$.
}
\label{fig:re_e_meta}
\end{figure*}

\subsubsection{Geometry}
Instead of handwritten-number-shaped scatterers, the testing scatterer here has the same shapes as objects in EMNIST. 
In the testing sample, the contrast of the c-shaped scatterer is unknown. 
$\mathbf{E}^\text{sca, Pred}$ is also investigated on two circles with radii of 3 m and 10 m in the $xoz$-plane, each having $N_r=360$ uniformly distributed observation points.
The REs of the predicted results are given in Table~\ref{tab:di_test31}. 
Figures~\ref{fig:pred_result_ds} and~\ref{fig:pred_result_3d_es_emnist} visualize $\mathbf{E}^\text{tot, Pred}$ and $\mathbf{E}^\text{sca, Pred}$. %
It can be seen that the proposed DL-based scheme delivers quite well generalization ability. 

\begin{figure*}[!tpb]
  \centering
  \includegraphics[width=0.8\textwidth]{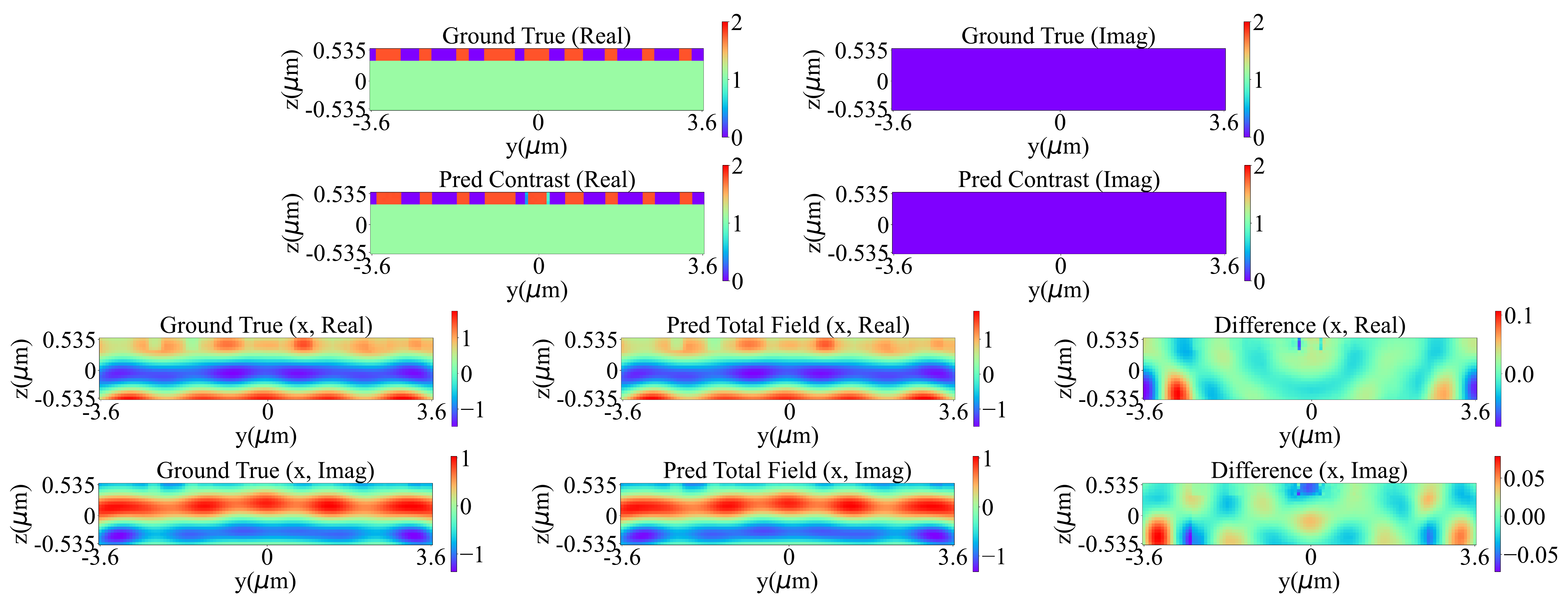}
\caption{Computations on the metasurface given in Fig.~\ref{fig:rel_error_es_center}: predicted results in the $yoz$-plane.
The real and imaginary parts of $\bm{\chi}^\text{Label}$ as well as $\bm{\chi}^\text{Pred}$ are shown in the first and second rows, respectively.
The real and imaginary parts of $\mathbf{E}^\text{tot}_x$ are in the third and forth rows. %
}
\label{fig:pred_result_metasurface}
\end{figure*}

\subsection{Numerical Experiments on Metasurfaces}
\label{sec:ne2m}
To further demonstrate the applicability of the proposed DL-based scheme,
rectangular-shaped dielectric metasurfaces are employed.  
The training set in this subsection is extended from a single incident angle to a multi-angle configuration where $\theta$ ranges from $-45^\circ$ to $45^\circ$ with a step of $10^\circ$ and $\phi$ is fixed at $90^\circ$.
Since 1,500 metasurfaces per incident angle are constructed by varying length and width of each meta-atom whose height is fixed, $10\times 1,500=15,000$ training samples are obtained in total.
Assuming the incident wave is a $x$-polarized plane wave with the wavelength $\lambda = 1.55 \mu m$, 
the input $\mathbf{E}^\text{sca}_\text{0}$ is sampled on three circles in the $xoy$-, $xoz$- and $yoz$-planes, each circle having a radius of $12 \mu m$ and 64 uniformly distributed observation points. 
One metasurface sample is presented in Fig.~\ref{fig:metamaterial}, which contains nine aligned meta-atoms with different shapes. 
The fused silica ($\chi=1.10$) substrate of the metasuface is of size $7.2\mu m \times 0.8\mu m \times 1.07 \mu m$, centering at (0, 0, 0) m.
Each meta-atom is a silicon ($\chi=1.79$) nanoblock with the thicknesses of $0.27\mu m$.
A $12 \times 108 \times 20$ grid is employed.
Denoting the 9 nanoblocks by $b_1$, $b_2$, $\cdots$, $b_9$, from left to right, as shown in the figure, one of the them (namely, $b_2$ in Fig.~\ref{fig:metamaterial}) is taken as unknown $p2$. 

The training costs around 76 h and the prediction process can finish in around 0.01 s. %
It would take MoM around 2.6 s to solve Eq.~\eqref{eq:dis_j} when both $\bm{\chi}^{p1}$ and $\bm{\chi}^{p2}$ are available.
A metasurface that is not included in the training set is chosen as the testing sample, whose top view is given in Fig.~\ref{fig:testsample}. 
The testing metasurface has the similar configuration to that of the training samples.
Three scenarios are discussed, where $b_1$, $b_5$ and $b_9$ are, respectively, taken as $p2$. 
In each scenario, the testing is conducted with $\theta$ ranging from $-60^\circ$ to $60^\circ$ with a step of $1^\circ$ and $\phi$ fixed at $90^\circ$.
$\mathbf{E}^\text{sca, Pred}$ is obtained at $N_r = 360$ observation points that are uniformly located on a circle with a radius of 15 $\mu m$ in the $yoz$-plane.
Its corresponding RE as a function of testing $\theta$ for each computation is given in Fig.~\ref{fig:re_e_meta}.
From the figure, it can be observed that when the incident wave comes from the directions where $p2$ is not sufficiently occluded, the proposed DL-based scheme exhibits smaller prediction errors, which aligns with what observed in Section~\ref{sec:sdia}.
Taking the computation shown in Fig.~\ref{fig:rel_error_es_center} as an example, when $p2$ is positioned in the center region, the scheme achieves smaller prediction errors across $\theta \in[-50^\circ, 50^\circ]$. 

%
\begin{figure}[!tpb]
  \centering
  \includegraphics[width=0.48\textwidth]{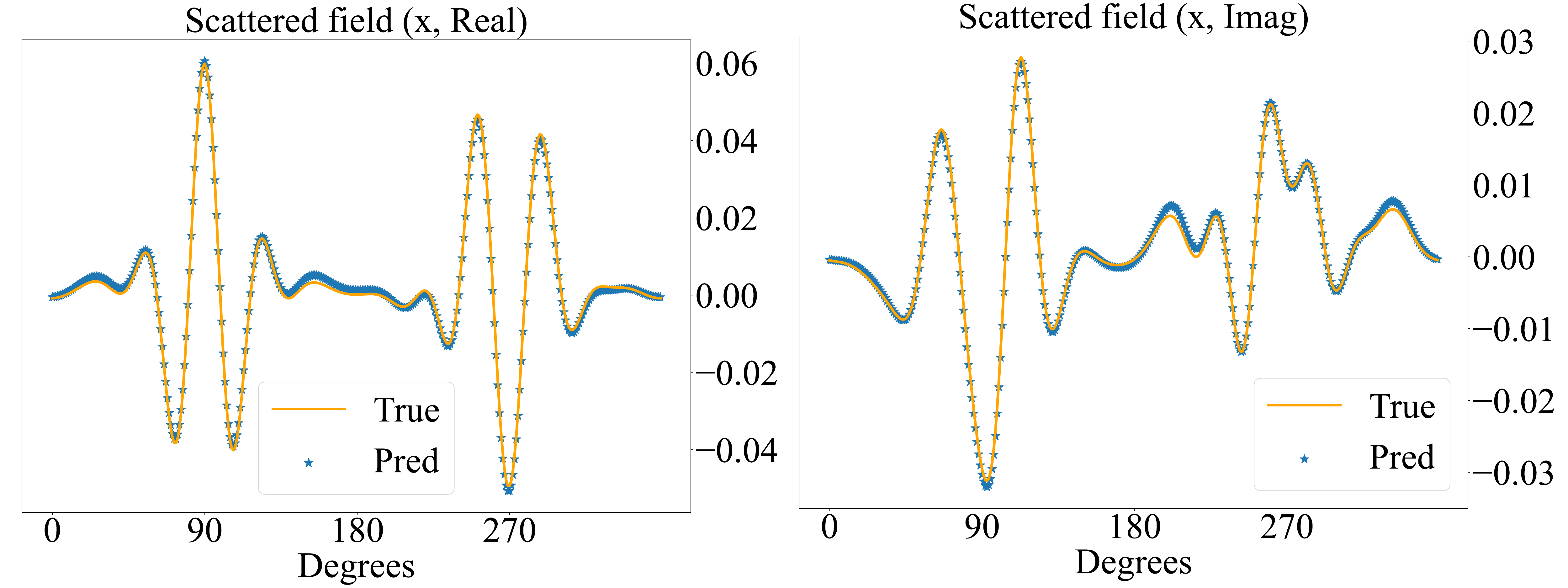}
\caption{Computations on the metasurface given in Fig.~\ref{fig:rel_error_es_center}: predicted $\mathbf{E}^\text{sca, Pred}_x$ in the $yoz$-plane ([$0^\circ$, $360^\circ$]). 
The first and second columns give the real and imaginary part of $\mathbf{E}^\text{sca, Pred}_x$, respectively.
}
\label{fig:pred_result_3d_es_meta}
\end{figure}

\begin{table}[!t]
  \begin{center}
    \caption{REs for computations presented in Figs.~\ref{fig:pred_result_metasurface} and~\ref{fig:pred_result_3d_es_meta}.
    }
    \label{tab:di_meta1}
    \setlength{\tabcolsep}{8mm}
    \renewcommand\arraystretch{1.3}
    \begin{tabular}{|cc|}
      \hline
      \multicolumn{1}{|c|}{$\bm{\chi}^\text{Pred}$ (\%)} & 2.94   \\ \hline
      \multicolumn{1}{|c|}{$\mathbf{E}^\text{tot, Pred}$ (\%)}    & 3.06   \\ \hline
      \multicolumn{1}{|c|}{$\mathbf{E}^\text{sca, Pred}$ (15 $\mu m$, \%)}    & 5.55  \\ \hline
      \end{tabular}
  \end{center}
\end{table}

Similarly, since $\mathbf{E}^\text{sca}_x$ is about $10^2$ larger than its $y$- and $z$-components in magnitude, Figs.~\ref{fig:pred_result_metasurface} and~\ref{fig:pred_result_3d_es_meta} only visualize $\mathbf{E}^\text{tot, Pred}_x$ and $\mathbf{E}^\text{sca, Pred}_x$ for the metasurface given in Fig.~\ref{fig:rel_error_es_center}. 
The incident direction is set to $\theta=0^\circ$ and $\phi=90^\circ$ that does not appear in the training set.
 The testing errors are listed in Table~\ref{tab:di_meta1}.
It can be seen that the proposed DL-based scheme achieves good generalization performance. 

\section{Conclusions}
\label{sec:Conclusion}
A DL scheme of EM forward scattering from dielectric scatterers whose profiles are incompletely known is discussed.
The proposed DL-based scheme is designed according to the formulation of the EM forward scattering from an incompletely known scatterer. 
Numerical experiments demonstrate that the performance of the proposed DL-based scheme can be quite good.
The generalization study reveals that the proposed scheme has the potential to solve scattering problems where complete profile of the scatterer is not available. 
It is also revealed that the larger amount of observation data available, the better generalization ability of the DL scheme can be.
Future work should be done to reveal how many observation data are sufficient for a required prediction accuracy.   


\bibliographystyle{IEEEtran}
\bibliography{IEEEabrv, reference.bib}

\end{document}